\documentclass[aps,prd,floatfix,nofootinbib,superscriptaddress,twocolumn]{revtex4-1}
\usepackage{graphicx}  
\usepackage{dcolumn}   
\usepackage{bm}        
\usepackage{amssymb}   
\usepackage{amsmath}
\usepackage{comment}
\usepackage{graphicx}
\usepackage{color}
\usepackage{slashed}
\usepackage{hyperref}
\usepackage{soul}

\usepackage[braket, qm]{qcircuit}

\newcommand{\Eq}[1]{Eq.~(\ref{#1})}
\newcommand{\Eqs}[2]{Eqs.~(\ref{#1}-\ref{#2})}
\newcommand{\Fig}[1]{Fig.~(\ref{#1})}

\newcommand{\xbj}{x_\text{Bj}}

\begin{document}

\author{Niklas Mueller}
\email{nmueller@bnl.gov}
\affiliation{Physics Department, Brookhaven National Laboratory, Bldg. 510A, Upton, NY 11973, USA}
\author{Andrey Tarasov}
\email{tarasov.3@osu.edu}
\affiliation{Physics Department, Brookhaven National Laboratory, Bldg. 510A, Upton, NY 11973, USA}
\affiliation{Department of Physics, The Ohio State University, Columbus, OH 43210, USA}
\affiliation{Joint BNL-SBU Center for Frontiers in Nuclear Science (CFNS) at Stony Brook University, Stony Brook, NY 11794, USA}
\author{Raju Venugopalan}
\email{raju@bnl.gov}
\affiliation{Physics Department, Brookhaven National Laboratory, Bldg. 510A, Upton, NY 11973, USA}

\title{Deeply inelastic scattering structure functions on a hybrid quantum computer}
\date{\today}
\begin{abstract}
We outline a strategy to compute deeply inelastic scattering structure functions using a hybrid quantum computer. Our approach takes advantage of the representation of the fermion determinant in the QCD path integral as a quantum mechanical path integral over 0+1-dimensional fermionic and bosonic worldlines. The proper time evolution of these worldlines can be determined on a quantum computer. While extremely challenging in general, the problem simplifies in the Regge limit of QCD, where the interaction of the worldlines with gauge fields is strongly localized in proper time and the corresponding quantum circuits can be written down. As a first application, we employ the Color Glass Condensate effective theory to construct the quantum algorithm for a simple dipole model of the $F_2$ structure function. We outline further how this computation scales up in complexity and extends in scope to other real-time correlation functions.  
\end{abstract}
\maketitle
\section{Introduction}
From its early role in the discovery of partons and asymptotic freedom,  deeply inelastic scattering (DIS) of electrons off protons and heavier nuclei has been a fundamental tool in studying the quark-gluon structure of matter~\cite{Breidenbach:1969kd,Bjorken:1968dy,Bjorken:1969ja,Gross:1973id,Blumlein:2012bf}. The simplest DIS quantities are the inclusive structure functions $F_{2}$ and $F_L$ which, respectively, provide information on the sum of the quark and antiquark distributions, and the gluon distribution, in the target proton or nucleus~\cite{CooperSarkar:1997jk,Arneodo:1992wf}.

Computing structure functions from first principles is an outstanding problem in Quantum Chromodynamics (QCD) because they are proportional to nucleon/nuclear matrix elements of products of electromagnetic currents that are light-like separated in Minkowski spacetime. In contrast, Monte Carlo computations in lattice QCD are robust in Euclidean spacetime. The operator product expansion (OPE)~\cite{Sterman:1994ce} allows one to compute moments of structure functions in lattice QCD, but such computations are restricted to low moments~\cite{Winter:2017bfs}.  There are interesting alternative developments  whereby so-called quasi- or pseudo-pdfs are computed on the lattice~\cite{Ji:2013dva,Alexandrou:2015rja,Chen:2016utp,Radyushkin:2016hsy};  excellent reviews of the status of ``classical" computation of structure functions can be found in \cite{Lin:2017snn,Detmold:2019ghl}. 

Given the formidable challenge, it is worthwhile to ask whether simulations on a quantum computer can overcome the limitations of classical Monte Carlo approaches. An appropriate analogy is that of the sign problem at finite baryon chemical potential in QCD~\cite{Ortiz:2000gc,Alexandru:2019ozf}. In this Noisy-Intermediate-Scale-Quantum (NISQ) era~\cite{Preskill:2018}, a path forward is to identify simple but scalable  problems that are tractable and to explore their implementation on quantum devices~\cite{Lloyd:1996,cirac2012goals,hauke2012can,Jordan:2011ci,Jordan:2011ne,nuclearReview,Preskill:2018fag,Lamm:2019bik}.

We will outline in this paper a digitization strategy whereby progress can be made constructing the Hilbert space of a relativistic quantum field theory from `worldline/single particle' states.
We illustrate this approach, as a simple first step, for the fermion sector of the QCD path integral, by expressing the fermion determinant in the QCD effective action as a quantum mechanical ``worldline" path integral~\cite{Strassler:1992zr,DHoker:1995uyv,DHoker:1995aat,Mondragon:1995va,Mondragon:1995ab,Hernandez:2008db,JalilianMarian:1999xt,Schubert:2001he,Bastianelli:2006rx, Corradini:2015tik,Bastianelli:2013pta,Mueller:2017arw,Mueller:2017lzw,Mueller:2019gjj}  over fermionic and bosonic variables~\footnote{For a worldline approach to quantum computing in other contexts, we refer the reader to \cite{kocia2017discrete,kocia2017discreteA,kocia2017semiclassical,kocia2019non,kocia2018stationary}.}, which when interpreted in a Hamiltonian formulation,  suggests a novel digitization strategy of the underlying quantum field theory.  The primary objective of this manuscript is to motivate and spell out the conceptual foundations of this approach with the eventual goal being a computation on quantum hardware. 

 In Section II, we will outline the essential elements of this worldline formalism. We will then in 
Section III, apply it to discuss the quantum computation of $F_2$ in an effective field theory approach to the high energy ``Regge" limit of QCD. Section IV will identify the quantum circuits needed for the  simplest toy model computation. In Section V, we will discuss the gradual scaling of the computation in complexity to address quantum computation in DIS to high orders in perturbation theory in 
the Regge limit. In Section VI, we outline the digitization strategy for bosonic worldlines in the context of a relativistic scalar field theory. A detailed discussion of the corresponding quantum circuits and 
their implementation on quantum hardware will be discussed in a follow-up paper~\cite{NAR}. In the concluding Section VII, we will discuss the expansion in scope of this worldline/single particle framework to address scattering amplitudes and other real time correlation functions that are classically challenging in both perturbative and nonperturbative QCD. We will outline the significant practical and conceptual challenges facing realistic computations of QCD real time distributions in the NISQ era and beyond. The paper has two Appendices. In the first Appendix, we discuss the preparation of the initial state of the proton in the worldline formalism. In Appendix B, we discuss some 
details of the quantum circuit discussed in Section IV.

\section{Worldline representation of the effective action.}
The Euclidean QCD+QED effective action can be expressed as 
 \begin{align}\label{eq:WLeffectiveaction}
\Gamma_E[A;a] =\frac{1}{2}\text{Tr}\log \left( D^2 -\frac{ig}{4}F_{\mu\nu}[\gamma^\mu,\gamma^\nu]\right)\,,
\end{align}
where the $A_\mu$ represent the QCD gauge fields, $a^\mu$ the QED photon field, with $g$ and $e$ the respective $SU(N_c)$ and $U(1)$ gauge couplings. Further, we have $D^2=D_\mu D^\mu$, with the covariant derivative $D_\mu=\partial_\mu-igA_\mu-ie a_\mu$, $F_{\mu\nu}=\frac{i}{g}[D_\mu,D_\nu]$ is the field-strength tensor, and the operator trace is over position, as well as color and spin degrees of freedom~\footnote{We will consider only light fermions and for simplicity will not include a mass term explicitly in most of the computations.}.
The logarithm in \Eq{eq:WLeffectiveaction} can be written in integral form 
and a heat-kernel regularization of this expression \cite{Strassler:1992zr} allows one to evaluate the functional trace in a coherent state basis for the coordinate $\hat{x}_\mu |x\rangle = x_\mu  |x\rangle$ and the momentum $\hat{p}_\mu |p\rangle = p_\mu  |p\rangle$, giving $\text{tr}_x(\mathcal{O}) = \int d^4x\, \langle x |\mathcal{O} | x\rangle = \int d^4p \,\langle p |\mathcal{O} | p\rangle$.
The trace over the Dirac matrix structure may be expressed in a coherent state basis of fermionic creation-annihilation operators~\cite{Strassler:1992zr,DHoker:1995uyv,DHoker:1995aat} 
\begin{align}\label{eq:representationGamnma}
\hat{a}_1^\dagger &= \frac{1}{2}\left( \gamma_1 + i\gamma_3 \right)\,,  \qquad \hat{a}_1 = \frac{1}{2}\left( \gamma_1 - i\gamma_{3} \right)\
\nonumber\\
\hat{a}_2^\dagger& = \frac{1}{2}\left( \gamma_2 +  i\gamma_{4} \right)\,,\qquad \hat{a}_2 = \frac{1}{2}\left( \gamma_2 -  i\gamma_{4} \right)
\end{align}
of the Clifford algebra of Euclidean Dirac matrices, satisfying $[\gamma_\mu, \gamma_\nu]_+ = 2\delta_{\mu\nu}$ ($\mu=1,2,3,4$). The  fermionic coherent states  are defined by
\begin{align}\label{eq:coherentstatesdef}
\hat{a}_i | \theta \rangle = \theta_i |\theta\rangle\,,\qquad
\hat{a}_i^\dagger | \theta^* \rangle = \theta_i^*|\theta^* \rangle\,,
\end{align}
where $\theta_i,\theta_i^*$, with $i=1,2$, are Grassmann variables.
The spinor trace in this coherent state basis is given by the functional integral $\text{tr}_s\mathcal{O} \equiv i \int d^2 \theta \langle -\theta | \mathcal{O} | \theta\rangle$ with the normalization $\text{tr}_s\mathbb{I} =4 $ \cite{Ohnuki:1978jv}. The Majorana representation of these is
\begin{align}\label{eq:Majorana}
\psi_1 &= \frac{1}{\sqrt{2}}(\theta_1^* + \theta_1)\,,\qquad \psi_3 = \frac{-i}{\sqrt{2}}(\theta^*_1-\theta_1)\,,
\nonumber\\
\psi_2 &= \frac{1}{\sqrt{2}}(\theta_2^* + \theta_2)\,,\qquad \psi_4 = \frac{-i}{\sqrt{2}}(\theta^*_2-\theta_2)\,.
\end{align}

Likewise, as outlined in Appendix \ref{app:densitymatrixproton}, the trace for $SU(3)$ color may be expressed as the Grassmann integral \cite{Barducci:1980xk,DHoker:1995uyv,DHoker:1995aat} $\text{tr}_c \mathcal{O} = \int d^3 \lambda \langle -\lambda | \mathcal{O}| \lambda \rangle$ by introducing the coherent states, $\hat{c}_a | \lambda \rangle = \lambda_a | \lambda \rangle $, $\hat{c}_a^\dagger | \lambda^* \rangle = \lambda_a^* | \lambda^* \rangle$, where $a=1,2,3$. However for the purposes of this discussion, we shall keep the trace over color explicit.

Employing these coherent states, and analytically continuing to Minkowski space ($-\Gamma_E\rightarrow i\Gamma$), one obtains the quantum mechanical path integral~\cite{Strassler:1992zr,Mueller:2017arw,Mueller:2017lzw} 
 \begin{align}\label{eq:effectiveaction1}
{\Gamma}[A;a]  = -\frac{i}{2} \int\limits_0^\infty \frac{dT}{T} \,\text{tr}_c \int\limits_{{P}} \mathcal{D}x \mathcal{D}p \int\limits_{{AP}}\mathcal{D}\psi  \,   e^{iS[A;a]}\,,
 \end{align}
with the action
  \begin{align}
 S[A;a] &= \int d\tau \big( p_\mu \dot{x}^\mu + \frac{i}{2} \psi_\mu \dot{\psi}^\mu  - H[A;a]\big)\,,
 \end{align}
 and the worldline Hamiltonian given by 
 \begin{align}\label{eq:WLHamiltonian}
 H[A;a]&= P^2 +ig\psi^\mu  F_{\mu\nu}[A]\psi^\nu +ie\psi^\mu  F_{\mu\nu }[a]\psi^\nu\,.
 \end{align}
Here $P_\mu = p_\mu -g A_\mu (x) -ea_\mu(x)$, $A_\mu=A^a_\mu t^a$ and $F_{\mu\nu}=F^a_{\mu\nu} t^a$, with $t^a$ the $SU(3)$ generators in the fundamental representation.
Further, $P$ ($AP$) denote periodic (anti-periodic) boundary conditions for commuting (anti-commuting) variables.

\section{DIS in the Regge limit}
In the inclusive DIS process $\ell(l) + N (P) \rightarrow \ell(l')+X$, the cross-section for the interaction between the lepton $(\ell)$ and the hadron $(N)$ can be factorized into the convolution of the lepton tensor $L_{\mu\nu}$, corresponding to the exchange of a virtual photon $\gamma^*$ with spacelike four-momentum $q=l-l' \equiv (q^-,q^+,0,0)$, and the hadron tensor $W^{\mu\nu}$ representing the interaction of the virtual photon with the parton constituents of the hadron~\cite{Peskin:1995ev}. The latter is given by the imaginary part of the forward Compton amplitude:
\begin{align}\label{eq:Tmunu}
W^{\mu\nu}(q,P,S)= \text{Im} \, \frac{i}{\pi}&  \int d^4\mathbf{x} \,e^{i\mathbf{q}\cdot \mathbf{x}} 
\langle P,S| \mathbb{T} \, \hat{j}^\mu(\mathbf{x}) \hat{j}^\nu(0) | P,S\rangle\,,\nonumber\\
\end{align}
where ``$\mathbb{T}$" denotes time ordering of the bilinear product of electromagnetic current operators $\hat{j}^\mu = \hat{{\bar \Psi}}\gamma^\mu \hat{\Psi}$ of quark fields $\hat{\Psi}$ in the hadron state with four-momentum $P$ and spin $S$. Structure functions extracted from $W_{\mu\nu}(q,P,S)$ are expressed in terms of the Lorentz scalars $Q^2=-q^2 >0$ and  $x_\text{Bj}=Q^2/(2P\cdot q)$. In particular, $F_2(x_\text{Bj},Q^2)$ is obtained from the projection~\footnote{Similar expressions for 
$F_1$ and $g_1,g_2$ can be found in \cite{Leader:2001gr}.} $F_2 \equiv \Pi_2^{\mu\nu} W_{\mu\nu}$, with $\Pi_2^{\mu\nu}\equiv \frac{3P\cdot q}{4a}[\frac{P^\mu P^\nu}{a}-\frac{g^{\mu\nu}}{3}]$, $a=P\cdot q/(2 \xbj)+M^2$, and $M$ the hadron mass.

In lattice gauge theory, direct evaluation of the time ordered product on the r.h.s of Eq.~(\ref{eq:Tmunu}) requires computing the ratio of four-point and two point Euclidean correlators and subsequent analytic continuation  from Euclidean to Minkowski space~\cite{Liu:1993cv,Aglietti:1998mz,Hansen:2017mnd,Liang:2019frk}. In section~\ref{app:current-current} (some details are also given in Appendix \ref{app:densitymatrixproton}), we outline a first principles computation of $W_{\mu\nu}$ in the worldline formalism which illustrates the complexity of the problem. We will return to these issues shortly. 

We will argue here that significant progress towards quantum computation of structure functions can be made in Regge kinematics, corresponding to $Q^2=$ fixed, and $x_{\rm Bj}\approx Q^2/s \rightarrow 0$~\footnote{Interestingly, these asymptotics are the most daunting for classical computing approaches since  the OPE breaks down at small $x_{\rm Bj}$~\cite{Mueller:1996hm}.}, where the squared center-of-mass energy $s\approx 2 P^+ q^-$. In this limit, a Born-Oppenheimer separation of scales appears between fast ($x_{\rm Bj}\sim 1$) and slow ($x_{\rm Bj} \ll1$) degrees of freedom~\cite{McLerran:1993ni,McLerran:1993ka,McLerran:1994vd}, allowing the former to be described as static color sources and the latter as dynamical gauge fields coupled to the sources.

This argument is quantified in the Color Glass Condensate effective field theory (CGC EFT)~\cite{Gelis:2010nm}, wherein an emergent scale  proportional to the density of color sources grows through a Wilsonian renormalization group evolution of the separation between sources and fields with decreasing $x_{\rm Bj}$~\cite{JalilianMarian:1997gr,JalilianMarian:1997dw,Iancu:2000hn,Ferreiro:2001qy}. In Regge asymptotics, this scale is larger than intrinsic nonperturbative QCD scales and therefore the hadron tensor in the CGC EFT can be written as~\cite{McLerran:1998nk,Tarasov:2019rfp},
\begin{align}\label{eq:hadrontensor}
&W^{\mu\nu}(q,P,S)= \frac{P^+}{\pi e^2} \text{Im} \int d^2 X_\perp \int dX^- \int_{k,k'}  \int d^4\mathbf{x} \nonumber \\
&\times e^{i{q}\cdot {x}} e^{-i k(X^- +\frac{x}{2})} e^{-i k'(X^- -\frac{x}{2})}
 \nonumber\\
 &\times\int [\mathcal{D}\rho] \, W[\rho]\int [\mathcal{D}A]  \, \tilde{\Gamma}^{\mu\nu}[k,k';A] \, e^{i\Gamma[A] + iS [A,\rho]}\,,
\end{align}
where $\int_{k} = \int d^4k /{(2\pi)^4}$.
We work in lightcone coordinates $x^\pm \equiv (x^0 \pm x^3)/\sqrt{2}$, $p^\pm \equiv(p^0 \pm p^3)/\sqrt{2}$, assuming a right moving hadron with large $P^+$.
In \Eq{eq:hadrontensor}, $S [A,\rho]=-\frac{1}{4}F^c_{\mu\nu}F^{c,\mu\nu} + J\cdot A$~\footnote{Note that the term $J\cdot A$ can also be written in a gauge invariant generalization~\cite{JalilianMarian:2000ad} but for the problem of interest here, this simpler form will suffice.}. Here $J^{\mu,c}{=} \,\delta^{\mu +}\rho^c(x^-,x_\perp)$, where the static large $x_{\rm Bj}$ color source density $\rho^c(x^-,x_\perp)$ ($c=1,\cdots, 8$) has support limited to $\Delta x^-=1/P^+$ and $W[\rho]$ is a gauge invariant weight functional representing the nonperturbative distribution of these sources. 
The polarization tensor $\tilde{\Gamma}^{\mu\nu}$ is given by 
\begin{align}\label{eq:poltensor}
i\tilde{\Gamma}^{\mu\nu}[k,k']=\int d^4z d^4z' \frac{i\delta \Gamma[A;a]}{\delta a_\mu(z) \delta a_\nu(z')}\Big|_{a=0} e^{ik\cdot z+ik' \cdot z'}\,,
\end{align}
where $ \Gamma[A;a]$ is the QCD+QED effective action with the worldline representation given in Eq.~(\ref{eq:effectiveaction1}). 
%
%
%

We will now discuss the computation
of $F_2$ on a quantum computer. The simplest problem we can address in 3+1-dimensions is to determine the quantum algorithm for the static ``shock wave" solution $A_{\rm cl}^\mu = (0,A_{\rm cl}^+,0,0)$ to the Yang-Mills equations in the CGC EFT where $A_{\rm cl}^{+,c}(x)= {\tilde \rho}^c (x_\perp)\delta(x^-)$, with $\rho^c(x_\perp,x^-) \approx {\tilde \rho}(x_\perp) \delta(x^-)$.  In this background, the worldline effective action can be written as 
\begin{align}\label{eq:effectiveaction}
&\Gamma [A_{\rm cl},a]=- \frac{i}{2} \text{tr}_c \int\limits_0^\infty \frac{dT}{T} \int d^4x  d^2 \theta  \langle x,-\theta| e^{-i\hat{H}[A_{\rm cl}, a]T} | x,\theta \rangle \,,
\end{align}
where the Hamiltonian operator $\hat{H}$ is obtained by quantizing \Eq{eq:WLHamiltonian}.  While consistent quantization requires that we eliminate states with indefinite metric from the physical subspace of the theory~\cite{Berezin:1976eg}, it will not be relevant for the computation of $F_2$ we consider here. We will return to this important issue in section~\ref{app:current-current}.

Computing the hadron tensor \Eq{eq:hadrontensor} from \Eq{eq:poltensor} and \Eq{eq:effectiveaction} facilitates the simplest possible hybrid quantum computation where only the spinor trace is simulated on a digital quantum computer. 
Starting from the worldline representation  of the effective action  in the 
shock wave background field $A_{\rm cl}$, given by \Eq{eq:effectiveaction}
\begin{figure}[tb]
\begin{center}
\includegraphics[width=0.32\textwidth]{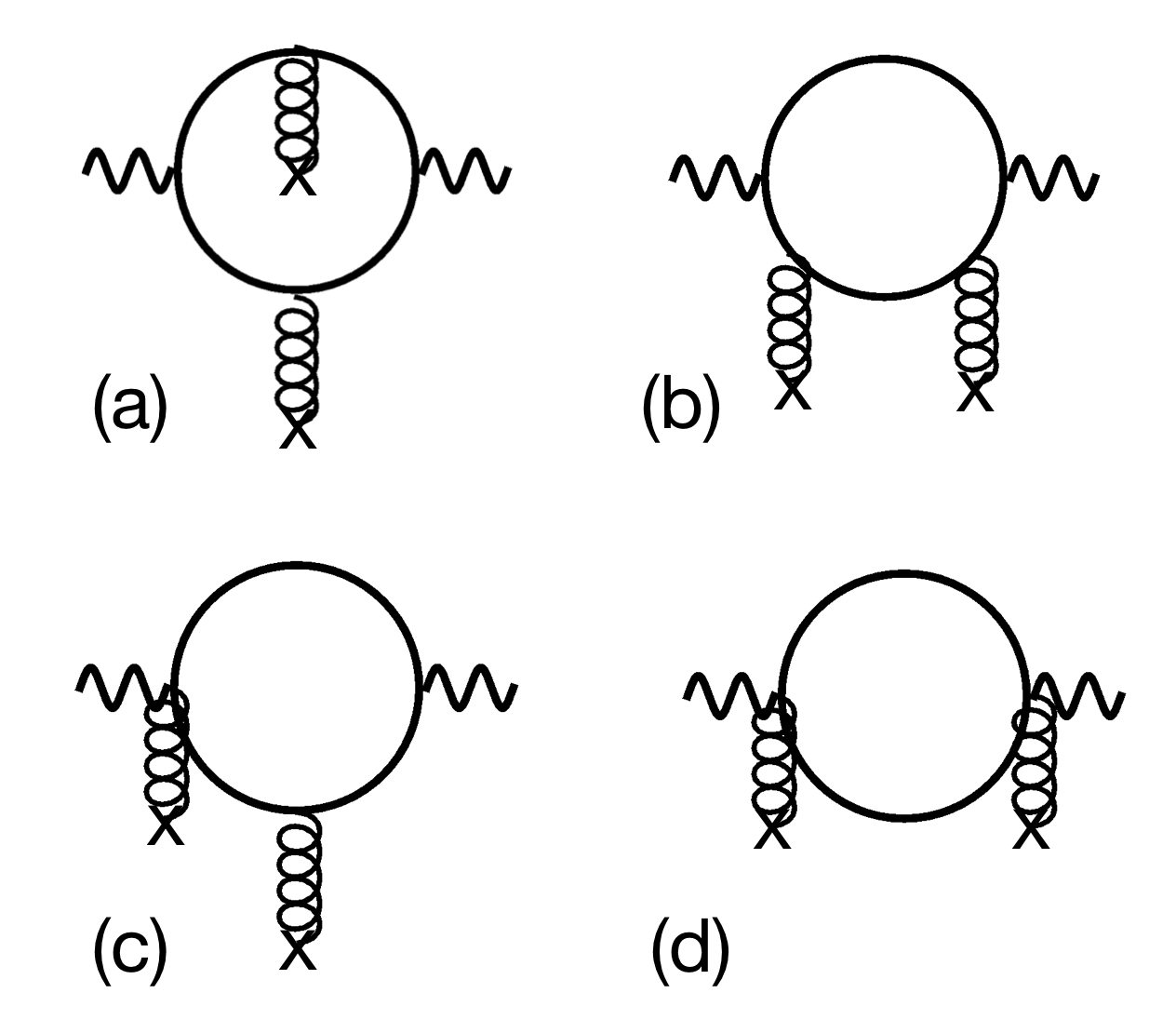}
\caption{Diagrammatic contributions to the photon polarization tensor in the hybrid worldline formalism.\label{fig:selfenergytopo}}
\end{center}
\end{figure}
where the worldline Hamiltonian $\hat{H}[A_{\rm cl}, a]$ is defined by Eq.~(\ref{eq:WLHamiltonian}) and depends on both the external electromagnetic field $a_\mu(x)$ and background gluon field of the target $A_\mu(x)$.  In the Regge limit of QCD, the imaginary part of the polarization tensor $\Gamma^{\mu\nu}$ in \Eq{eq:poltensor} has the physical interpretation of the virtual photon (emitted by the electron) splitting into a color singlet $q\bar{q}$ ``dipole"  long before its interaction with the target. This interaction, which has the structure of a shock wave,  is instantaneous and localized on the world-line at two arbitrary instants $\tau$ and $\tau'$ in proper time.

The quantum computation of the trace over the fermionic degrees of freedom in $\Gamma$, given by the integral $\int d^2 \theta$, is our principal objective here. As discussed in 
Appendix~\ref{app:densitymatrixproton}, the trace over color can also be expressed as a quantum mechanical path integral in the worldline action. For the purposes of this computation, to keep things simple, we will keep the trace over color explicit. 

We first divide the worldline effective action into $N$ segments of size $\delta\tau$. Since the interaction $A_{\rm cl}$ is localized, at most two segments of the worldline can interact with the background field. As a result, we get 
\begin{align}\label{eq:efeqsegments}
&\Gamma [A_{\rm cl},a] = -\frac{i}{2}  \text{tr}_c\int \limits_0^\infty \frac{dT}{T} \int d^4x\int d^2\theta \sum^{N}_{\substack{i,j = 1\\ i<j}}
\langle x ,-\theta | \hat{S}_{N, j+1}(a) \nonumber \\
&\times e^{-i\hat{H}_j[A_{\rm cl}, a = 0]\delta\tau} \hat{S}_{j-1,i+1}(a) e^{-i\hat{H}_i[A_{\rm cl}, a = 0]\delta\tau} \hat{S}_{i-1, 1}(a)  |x ,\theta \rangle\,.
\end{align}
Here  $\hat{H}_i[A_{\rm cl}, a = 0]$ is the Hamiltonian of segment number $i$ where the interaction with the shock wave background field is located and where the photon field $a_\mu(x) = 0$. Further, $\hat{S}_{m, n}(a)$ denotes segments where the converse is true; the worldline only interacts with the external photon field:
\begin{eqnarray}\label{eq:Sdef}
 \hat{S}_{m, n}(a) = \prod\limits_{i=n}^m  e^{-i\hat{H}_i[A = 0, a]\delta\tau}
 \end{eqnarray}
 and $ \hat{S}_{m, n}(a) = 1 $ if $m<n$.

In general, there could be a segment of the worldline containing both the shock wave $A_{\rm cl}$ and the photon field $a_\mu$ defined by the general operator $\exp{(-i\hat{H}_n[A, a]\delta\tau)}$. Examples of these segments are shown in \Fig{fig:selfenergytopo}(c) and (d). Indeed, there are several contributions of this type in \Eq{eq:poltensor} and one has to sum over all of these contributions. However in the computation of $F_2$, there is a cancellation between terms with the structure of  Figs.~(\ref{fig:selfenergytopo})(c) and (\ref{fig:selfenergytopo})(d). As a result, only the diagrams in \Fig{fig:selfenergytopo}(a) and \Fig{fig:selfenergytopo}(b) contribute.
To discuss the contribution of \Fig{fig:selfenergytopo}(a) to Eq. (\ref{eq:efeqsegments}), we insert momentum states $p_1$ and $p_2$ adjacent to the shock wave interaction segment (labelled r below), using the completeness relation
\begin{align}
\int d^4p_{1,2} |p_{1,2}\rangle \langle p_{1,2}| = 1\,,
\end{align}
yielding
\begin{align}
&\langle p_1|  e^{-i\hat{H}_r[A = 0, a]\delta\tau} |p_2 \rangle= i e \langle p_1| (\hat{p} \hat{a} + \hat{a} \hat{p}
 - 2 i \hat{\psi}^\mu \hat{\psi}^\nu \partial_\mu \hat{a}_\nu )\delta\tau |p_2 \rangle\,.
\label{eq:photint}
\end{align}
The r.h.s here is obtained by expanding the evolution operator in Eq. (\ref{eq:efeqsegments}) up to linear terms in the photon field $a_\mu(x)$. (Since there are no contact terms in  \Fig{fig:selfenergytopo}(a), it is safe to omit quadratic terms in the expansion).

Next inserting first a complete set of coordinate states, and subsequently taking the Fourier transformation of the first derivative of Eq. (\ref{eq:photint}) with respect to the external photon field, we obtain the relation
\begin{align}
&\int d^4z \frac{\delta }{\delta a_\mu(z)} \langle p_1| e^{-i\hat{H}[A=0, a]\delta\tau} |p_2\rangle e^{ik z} 
\label{eq:intvphoton}\\
&= ie\int d^4z \int d^4x \frac{\delta }{\delta a_\mu(z)} \langle p_1| x\rangle (p_1 a(x) + a(x) p_2
\nonumber\\
&~~~~~~~~~~~~~ - 2 i \hat{\psi}^\rho \hat{\psi}^\sigma \partial_\rho a_\sigma(x) )  \langle x |p_2\rangle e^{ik z} \delta\tau
\nonumber\\
&= ie \int d^4x \langle p_1| x \rangle \Big[ (p^\mu_1 + p^\mu_2) - 2 \hat{\psi}^\mu \hat{\psi}^\rho k_\rho \Big] \langle x |p_2 \rangle e^{ik x} \delta\tau
\nonumber
 \end{align}
which gives the structure of the interaction segment of the worldline with the external photon of momentum $k$.

Similarly, one can derive the form of the interaction segment between the worldline and the shock wave:
\begin{align}
&\langle p_1| e^{-i\hat{H}[A, a= 0]\delta\tau} |p_2\rangle = - \int d^4x  \int \frac{dk^+}{2\pi} ~ e^{- ik^+ x^-}~ 
\nonumber\\
&\times \langle p_1| x \rangle \Big[ (p^-_1 + p^-_2) + 2 i  \hat{\psi}^- \hat{\psi}^m \partial_m  \Big]
\nonumber\\
&\times e^{ - i g \int A_{\rm cl}^+(z^-, \, x_{\perp}) dz^-} \langle x |p_2 \rangle \delta \tau\,,
\label{eq:intvsw}
\end{align}
where the phase in $A_{\rm cl}^+$ denotes multiple scattering off the background field. 
Substituting the explicit form of the interaction vertices  Eqs.~(\ref{eq:intvphoton}) and (\ref{eq:intvsw}) into Eq.~(\ref{eq:efeqsegments}), yields for the contribution from \Fig{fig:selfenergytopo}(a), 
\begin{widetext}
\begin{align}
\label{eq:semifinalresult}
&W_{(a)}^{\mu\nu}(q, P)= - \frac{ P^+ \sigma }{16\pi q^-}  \text{tr}_c\,\int d^2x_\perp \int \frac{d^2k_\perp}{(2\pi)^3} e^{ik_\perp x_\perp } \int^1_0 dz \int \frac{d^2p_\perp}{(2\pi)^2}
 \frac{1}{ \{ p^2_\perp + m^2 + z (1 - z) Q^2 \} \{ (p - k)^2_\perp  + m^2 + z (1 - z )Q^2 \}}
 \nonumber\\
&\times \Big\{ i \int d^2\theta_0 \langle -\theta_0 | \Big[ (2p^- - 2q^- - k^- ) + 2  \hat{\psi}^- \hat{\psi}^m k_m \Big] e^{- i g \int^\infty_{-\infty} A_{\rm cl}^+(x^-, x_\perp) dx^-} \Big[ ( 2p^\mu - q^\mu ) + 2 \hat{\psi}^\mu \hat{\psi}^\rho q_\rho \Big]
\nonumber\\
&\times\Big[ ( 2p^- - k^- )  -  2 \hat{\psi}^- \hat{\psi}^n k_n \Big] e^{- i g \int^{-\infty}_\infty A_{\rm cl}^+(x^-, 0_\perp) dx^-}\Big[ (2p^\nu - q^\nu - 2k^\nu ) - 2 \hat{\psi}^\nu \hat{\psi}^\rho q_\rho \Big] | \theta_0 \rangle \Big\}\Big|_{k^- = 0}\,,
\end{align}
where we integrated over intermediate coordinates and momenta, as well as over the positions of the interaction segments on the worldline. Note that we introduced the variable $z = p^-/q^-$. The diagram in \Fig{fig:selfenergytopo}(b) has a similar structure allowing one to compute $W_{(b)}^{\mu\nu}(q, P)$. Further technical details of the worldline computation of this  ``dipole model"  can be found in Ref. \cite{Tarasov:2019rfp}. 

To compute $F_2$, we need to sum both contributions and consider the projection
\begin{align}
&F_2 = \frac{Q^2}{4P\cdot q}\Big[\frac{3Q^2}{(P\cdot q)^2}P^\mu P^\nu - g^{\mu\nu}\Big] W_{\mu\nu}\,,
\end{align}
where $W_{\mu\nu}=W^{(a)}_{\mu\nu}+W^{(b)}_{\mu\nu}$. After integrating over transverse momenta $p_\perp$ and $k_\perp$ and summing over flavors $f$, we get our final result
\begin{align}
&F_2(q, P) =  \frac{   Q^2 \sigma N_c }{ 16\pi^3 }\sum\limits_f  e_f^2\, \int^1_0 dz \int dx_\perp ~ x_\perp ~ \Big[  1 - \frac{1}{N_c}\text{tr}_c ~ \mathcal{U}_\rho( x_\perp) \mathcal{U}_\rho^\dag( 0_\perp) \Big] 
\nonumber\\
&\times ~ i \int d^2\theta_0 \langle -\theta_0 |  - 3Q^2 z ( 1 - z ) [ ( 2z - 1 ) + 2 \hat{\psi}^- \hat{\psi}^+ ] [ (2z - 1 ) - 2 \hat{\psi}^- \hat{\psi}^+ ] K^2_0(\bar{Q}_f^2 x_\perp)
\nonumber\\
&+ 2 \{ z^2 + (1-z)^2 \} \bar{Q}_f^2 K^2_1(\bar{Q}_f^2 x_\perp) - 4 z ( 1 - z ) Q^2\hat{\psi}^- \hat{\psi}^+   \hat{\psi}^- \hat{\psi}^+ K^2_0(\bar{Q}_f^2 x_\perp)
\nonumber\\
& -  4 z ( 1 - z ) Q^2 \hat{\psi}^j \hat{\psi}^+ \hat{\psi}^j \hat{\psi}^- K^2_0(\bar{Q}_f^2 x_\perp) + z ( 1 - z ) ( 4 z^2 - 4 z + 3 ) Q^2 K^2_0(\bar{Q}_f^2 x_\perp)  + 2 m^2_f K^2_0(\bar{Q}_f^2 x_\perp) | \theta_0 \rangle  \,
\end{align}
\end{widetext}
where $\bar{Q}_f^2 = z(1-z)Q^2 + m_f^2$ and
\begin{align}\label{eq:SW2}
&\mathcal{U}_\rho(x_\perp)= \exp{ \Big\{-ig\int_{-\Delta x^-/2}^{\Delta x^-/2} dx^- \, A_{\rm cl}^+(x^-,x_\perp) \Big\}}
\end{align}
where $\Delta x^- = 1/P^+ \rightarrow 0$ is the width of the shock wave in the high energy limit.  
The $x_\perp$ independent terms stem from the diagram in \Fig{fig:selfenergytopo}(b) and ensure the UV finiteness of the expression at this order. Finally combining the terms containing the MacDonald functions $K_{0,1}$ into the well-known $\gamma^\star \rightarrow q\bar{q}$ wave functions $\Psi^f_{L,T}$, 
we arrive at \footnote{As noted earlier, $F_L$ can be obtained by taking a different kinematic projection of $W^{\mu\nu}$.}
\begin{align}\label{eq:F2}
&F_2(q,P)=
\frac{\sigma \, Q^2 }{2\pi e^2} \int [\mathcal{D} \rho] W[\rho]  \int\limits_{x_\perp} \int\limits_z  \sum\limits_{L,T;\, f} |\Psi^f_{L,T}(z, x_\perp)|^2
\nonumber\\
&\times D_\rho(x_\perp) \, i \int d^2\theta \langle -\theta | \big[ {\Omega}_{L,T}(z, x_\perp)\big] | \theta \rangle\,.
\end{align}
Here $ \int_{x_\perp}\equiv \int d^2 x_\perp$, $\int_z\equiv \int_0^1 dz$,  $\sum_{L,T; f}$ is the sum over the photon polarization and quark flavors,  $\sigma = \int d^2 X_\perp$ is the transverse radius of the hadron/nucleus and $ |\Psi_{L/T}(x_\perp,z)|^2$ denotes the modulus squared of the wave function of a virtual photon with longitudinal (L) or transverse (T) polarization to split into a quark-antiquark ``dipole",
\begin{align}
|\Psi^f_L(z, {x}_\perp)|^2&= \frac{e^2_f e^2N_c}{2\pi^3} Q^2 z^2(1-z)^2 \,K_0^2(\bar{Q}_f {x}_\perp)\,,\label{eq:photonwavefunctionL}\\
|\Psi_T^f(z, x_\perp)|^2&=\frac{e_f^2 e^2 N_c}{8\pi^3}~\big[\big(z^2 + (1-z)^2\big) \bar{Q}^2_f K^2_1(\bar{Q}_f x_\perp)
\nonumber\\
&+m_f^2 K^2_0(\bar{Q}_f x_\perp)\big]\label{eq:photonwavefunctionT}\,,
\end{align}
where $e_f$ represents the fractional charge of a quark of flavor $f$, $\bar{Q}_f^2\equiv z(1-z)Q^2 + m_f^2$. The dipole-hadron cross-section is given by $D_\rho(x_\perp) = \frac{1}{N_c} tr_c  \big[{1- \mathcal{U}}_\rho(x_\perp)  { \mathcal{U}}_\rho^\dagger(0_\perp)\big]$ for a given $\rho$.

In \Eq{eq:F2}, $\Omega_{L,T}$ are given by
\begin{align}\label{eq:omega}
&\Omega_L(z,x_\perp)= \frac{1}{ 2z(1-z) } \Big\{ - \frac{3}{4} ~ [ ( 2z - 1 ) + 2 \hat{\psi}^- \hat{\psi}^+ ] 
\nonumber\\
&\times[ (2z - 1 ) - 2 \hat{\psi}^- \hat{\psi}^+ ] - \hat{\psi}^+ \hat{\psi}^-   \hat{\psi}^+ \hat{\psi}^-  -   \hat{\psi}^j \hat{\psi}^+ \hat{\psi}^j \hat{\psi}^- 
\nonumber\\
&- z(1 - z) + \frac{3}{4} \Big\}\,,
\end{align}
where $\Omega_T(z,x_\perp)=1$ is trivial and need not be quantum computed, while 
$[2z-1\pm2{\psi}^- {\psi}^+]$ and $\psi^j\psi^\pm $ are the photon vertex insertions described further below. We  note finally that  we set $\exp(i\Gamma)$ in Eq.~(\ref{eq:hadrontensor}) to unity, which is valid to leading order in the coupling.

%
%
%
%

\section{Worldline Algorithm for the dipole model.}
The trace in \Eq{eq:F2} of \Eq{eq:omega} can be determined on a quantum
computer. We quantize $\psi^\mu\rightarrow \hat{\psi}^\mu=\gamma_5 \gamma^\mu/\sqrt{2}$, where $\gamma^\mu$ are the Dirac matrices in Minkowski spacetime satisfying $[\gamma_\mu, \gamma_\nu]_+ = 2g_{\mu\nu}$ with $(+,-,-,-)$ signature and $\gamma_5=i\gamma^0\gamma^1\gamma^2\gamma^3$. We then replace $\hat{\psi}^0 = (\hat{b}_1^\dagger -\hat{b}_1 )/\sqrt{2}$, $\hat{\psi}^3 = (\hat{b}_1^\dagger +\hat{b}_1 )/\sqrt{2}$, $\hat{\psi}^1 = (\hat{b}_2^\dagger +\hat{b}_2 )/\sqrt{2}$ and $\hat{\psi}^2 = -i(\hat{b}_2^\dagger -\hat{b}_2 )/\sqrt{2}$, where $\hat{b}_i^\dagger, \hat{b}_i$ ($i=1,2$) are fermion creation and annihilation operators satisfying $[\hat{b}_i^\dagger,\hat{b}_j]_+ = \delta_{ij}$. Further, we define lightcone operators $\hat{\psi}^+ = \hat{b}^\dagger_1$
and $\hat{\psi}^-=-\hat{b}_1$.

Performing a Jordan-Wigner transformation 
$\hat{b}_1^\dagger = (\sigma^x- i\sigma^y)/2\otimes \mathbb{I}$, $\hat{b}_1 = (\sigma^x+ i\sigma^y)/2\otimes \mathbb{I}$,  $\hat{b}_2^\dagger = \sigma^z\otimes (\sigma^x- i\sigma^y)/2$ and $\hat{b}_2 = \sigma^z\otimes (\sigma^x+ i\sigma^y)/2$
 we can write the individual terms\ in \Eq{eq:omega} and \Eq{eq:SW2} as
\begin{align}\label{eq:photonvertex}
\hat{\psi}^- \hat{\psi}^+ &= -\frac{1}{2}\,  [\mathbb{I}+ \sigma^z] \otimes \mathbb{I}\,,
\nonumber\\
\hat{\psi}^1 \hat{\psi}^\pm &= - \frac{1}{2\sqrt{2}} \, [\sigma^x\mp i \sigma^y] \otimes \sigma^x \,,
\nonumber\\
\hat{\psi}^2 \hat{\psi}^\pm &=  \frac{1}{2\sqrt{2}}\,  [\sigma^x\mp i \sigma^y] \otimes \sigma^y \,,
\end{align}
allowing us to express the shock wave operator and the photon vertex terms as quantum circuits of 2-qubit operations involving tensor products of the Pauli spin operators and the unit operator.

One can evaluate the spin traces in \Eq{eq:F2} with the quantum circuit
\begin{align} \label{eq;tracecircuit}
\Qcircuit @C=.5em @R=-.5em @!R {
&\lstick{ \hat{\rho}_c = | 0 \rangle \langle 0 |}   & \gate{H} &  \ctrl{1} & \qw &  \qw & \meterB{ \sigma }\inputgroup{2}{3}{.95em}{\hat{\rho}_n= \mathbb{I}_n /2^n } \\
& &  \lstick{  }& \multigate{1}{{\Omega}_{L,T}} & \qw  & \qw \\
 & & \lstick{ (n \text{ qubits})} & \ghost{{\Omega}_{L,R}} & \qw & \qw
}\,,
\end{align}
where $\hat{\rho}_c=|0 \rangle \langle 0|$ is an auxiliary control qubit $\hat{\rho}_c$ and $\hat{\rho}_n= \mathbb{I}_n /2^n $ initially, which combine to form the density matrix $\hat{\rho} = \hat{\rho}_c \otimes \hat{\rho}_n$ (valid for any unitary $n$-qubit operator ${\Omega}_{L,T}$)~\cite{knill1998power,datta2005entanglement,shepherd2006computation}.  The elements of the circuit, specified in detail in Appendix~\ref{app:circuit}, include the Hadamard gate $H$, the controlled-$\Omega_{L,T}$ ($C(\Omega_{L,T})$) gate, and the measurement of the Pauli operators $\langle \sigma^x \rangle $ and $\langle \sigma^y\rangle$. The action of the latter on the control qubit yields the real and imaginary part of $\text{Tr}[{\Omega}_{L,T}] $, respectively \cite{datta2005entanglement}. The controlled-$\Omega_{L,T}$ gate can be straightforwardly constructed and is given in Appendix~\ref{app:circuit}. This completes the quantum algorithm to measure the worldline trace in \Eq{eq:F2}.

%
%
%
\section{Expanding in complexity} \label{app:current-current}
We presented above a quantum circuit for the worldline representation of the fermion determinant in the limit of a localized proper time interaction.
The toy problem we outlined has the virtue that analytical results for $F_2$ are known for specific choices of $W[\rho]$  and therefore provide a benchmark to test our quantum algorithm. It can however be  expanded significantly in complexity. To appreciate this, we note that the r.h.s of \Eq{eq:Tmunu} is the ``in-in" matrix element of a real-time correlation function, where 
\begin{align}\label{eq:protonstateformalmaintext}
| P,S \rangle = \hat{\mathcal{U}}_{(0,-\infty)}\, \hat{\Phi}_{P,S} |0\rangle\,,
\end{align}
represents the state of the hadron (specifically a proton) before its interaction with the virtual photon. Here  $\hat{\mathcal{U}}_{(t,t')}\equiv \exp{(-i \hat{H}(t-t'))}$ where $\hat{H}$ is the QCD Hamiltonian. The operator $\hat{\Phi}_{P,S}$ creates a ``valence" quark and gluon state with the proton's quantum numbers from the non-interacting vacuum at past infinity, $\hat{\rho}_\text{init}\equiv \hat{\Phi}_{P,S} |0\rangle (\hat{\Phi}_{P,S} |0\rangle)^\dagger$; the proton is the result of its subsequent evolution with the QCD Hamiltonian. 

\begin{figure}[tb]
\begin{center}
\includegraphics[width=0.35\textwidth]{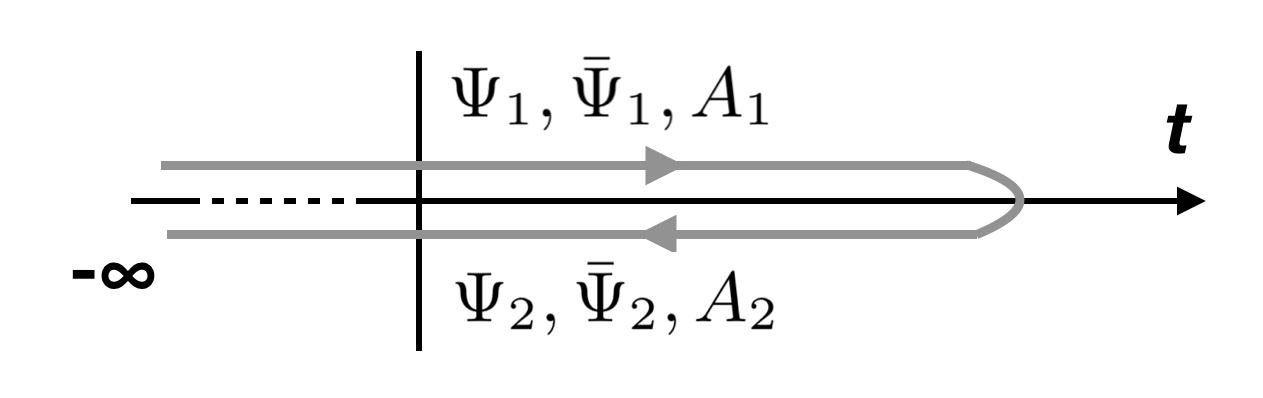}
\caption{Visualization of the Schwinger-Keldsyh contour \Eq{eq:SKcontourQCD}.\label{fig:SKcontourfields}}
\end{center}
\end{figure}

The proton state defined by \Eq{eq:protonstateformalmaintext} is encoded in a partition function $Z=\text{Tr} (\hat{\rho}_{PS})$, where  $ \hat{\rho}_{PS} = | P,S \rangle \langle P,S |$ is the proton's density matrix.
It is related to the initial state $\hat{\rho}_{\text{init}} \equiv \hat{\Phi}_{P,S} |0\rangle(\hat{\Phi}_{P,S} |0\rangle)^\dagger$ by the QCD evolution operator and may be expressed as the Schwinger-Keldysh path integral~\cite{Schwinger:1960qe,Keldysh:1964ud}
\begin{align}\label{eq:SKcontourQCD}
&Z=\text{Tr} [\hat{\rho}_{PS}]=\text{Tr}\left[ \hat{\mathcal{U}}_{(0,-\infty)} \,\hat{\rho}_{\text{init}}\, \hat{\mathcal{U}}_{(-\infty,0)}\right] = \int dA_1dA_2  
\nonumber\\
&\times  \int d \Psi_1d \Psi_2 \langle A_1, \Psi_1 | \hat{\rho}_{\text{init}} | A_2, \Psi_2\rangle\int\limits_{A_1}^{A_2} \mathcal{D}A \int\limits_{\Psi_1}^{\Psi_2} \mathcal{D} \Psi \mathcal{D} \bar{\Psi} \,e^{i S_\mathcal{C}}\,,
\end{align}
where $S_\mathcal{C}=\int d^4x_{\mathcal{C}}   \{   -\frac{1}{2} \text{tr}_c F_{\mu\nu}F^{\mu\nu} +   \bar{\Psi}(i \gamma^\mu D_\mu[A]+m) \Psi \}$ is the QCD action which has support on the Keldysh double time contour $\mathcal{C}$ depicted in \Fig{fig:SKcontourfields}. We abbreviated $D_\mu\equiv \partial_\mu -ig A_\mu$,  $dA_{1/2} \equiv \prod_{\mathbf{x},\mu} dA_{\mu,1/2}(\mathbf{x})$, where $A_{\mu,1/2}(\mathbf{x})\equiv A_\mu(\mathbf{x}_{1/2})$,
 $\mathcal{D}A\equiv \prod_{\mathbf{x},t_\mathcal{C},\mu}dA_\mu(\mathbf{x},t)$ and likewise for the fermionic integrals. 
 
 In \Eq{eq:SKcontourQCD}, $\langle A_1, \Psi_1 | \hat{\rho} | A_2, \Psi_2\rangle$ are matrix elements of an initial density matrix of non-interacting quarks and gluons at $t\rightarrow -\infty$,  $\hat{\rho}_{\text{init}} = \hat{\rho}_{\text{YM}} \otimes \hat{\rho}_V$, where $\hat{\rho}_{\text{YM}} = | 0 \rangle \langle 0|$ is the Yang-Mills vacuum and $\hat{\rho}_V$
is a three valence quark state with the proton's quantum numbers. Aiming at performing the fermionic path integral, we may write
\begin{align}
Z=  \int dA_1dA_2 \langle A_1 | \hat{\rho}_\text{YM} | A_2\rangle  \int\limits_{A_1}^{A_2}  \mathcal{D}A  \, Z_f[A] \, \exp{ \{ i S_\mathcal{C}^{\text{YM}} \}}\,,
\end{align}
where 
\begin{align}\label{eq:quarkpartQCD}
Z_f[A]\equiv \int  & d \Psi_1d \Psi_2 \langle \Psi_1 | \hat{\rho}_V |  \Psi_2\rangle
\int\limits_{\Psi_1}^{\Psi_2} \mathcal{D} \Psi \mathcal{D} \bar{\Psi} \,\exp {  \big\{ i S_{\mathcal{C}}^q \big\} } \,,
\end{align}
with $ i S_{\mathcal{C}}^q\equiv \int d^4z_{\mathcal{C}}   \bar{\Psi}(i  \slashed{D}[A]+m) \Psi $ and
$S^{\text{YM}}_{\mathcal{C}} \equiv -\frac{1}{2}\int d^4z_\mathcal{C} \, \text{tr}_c F^{\mu\nu}F_{\mu\nu}$.

Before considering the more complicated case of a baryon with three quarks, we consider first that of a single valence quark. We will also ignore the flavor
part of this single valence quark state.   Matrix elements for momentum, spin and color of its initial density matrix
 $\hat{\rho}_q = | \mathbf{p}, s, c \rangle \langle  \mathbf{p}, s, c |$  are given by 
\begin{align}\label{eq:singlequarkdensity}
\langle& \Psi_1 | \hat{\rho}_q |  \Psi_2\rangle=\langle \Psi_1 |   \mathbf{p}, s, c \rangle \langle  \mathbf{p}, s, c|  \Psi_2\rangle
\nonumber\\
& =2E_{\mathbf{p}}\, \int d^3\mathbf{x}_1d^3\mathbf{x}_2 [u^\dagger_{\mathbf{p},s, c} \Psi(\mathbf{x}_1)]
[\Psi^\dagger(\mathbf{x}_2)  u_{\mathbf{p},s, c}] e^{i\mathbf{p}(\mathbf{x}_1 - \mathbf{x}_2)}\,,
\end{align}
where $u_{\mathbf{p},s, c} \equiv u_{\mathbf{p},s} \otimes \Phi_c$
represents the quark spinor and color wave function, and $E_\mathbf{p} \approx|\mathbf{p}|$ for light quarks. The QCD path integral \Eq{eq:quarkpartQCD} may be written in terms of this initial condition as
\begin{align}
&Z_f[A]\equiv 
\int   d \Psi_1d \Psi_2 2E_\mathbf{p}\, \int d^3\mathbf{x}_1d^3\mathbf{x}_2\,[u^\dagger_{\mathbf{p},s, c} \cdot \Psi(\mathbf{x}_1)]
\nonumber\\
&\times[\Psi^\dagger(\mathbf{x}_2) \cdot  u_{\mathbf{p},s, c}] e^{i\mathbf{p}(\mathbf{x}_1 - \mathbf{x}_2)}\int\limits_{\Psi_1}^{\Psi_2} \mathcal{D} \Psi \mathcal{D} \bar{\Psi}\,\exp {  \big\{ i S_{\mathcal{C}}^q \big\} } 
\nonumber\\
&=2E_\mathbf{p}\, \int d^3\mathbf{x}_1d^3\mathbf{x}_2 \, e^{i\mathbf{p}(\mathbf{x}_1 - \mathbf{x}_2)}\big[{u}^\dagger_{\mathbf{p},s, c}\big]_i \,  \big[ \gamma^0 u_{\mathbf{p},s, c}\big]_j
\nonumber\\
 &\times \frac{\delta^2}{\delta \bar{J}_i(\mathbf{x}_1) \,\delta J_j(\mathbf{x}_2)}  \int \mathcal{D} \Psi' \mathcal{D} \bar{\Psi}' \, e^{ iS_{\mathcal{C}}^q + i\int d^4x_\mathcal{C} (\bar{J} \Psi + \bar{\Psi} J  ) }\Big|_{J=\bar{J}=0}
 \nonumber\\
 &=\det[ -i \hat{G}^{-1} ]\, 2E_\mathbf{p} \int d^3\mathbf{x}_1d^3\mathbf{x}_2  \, e^{i\mathbf{p}(\mathbf{x}_1 - \mathbf{x}_2) }
  \nonumber\\
 &\times \text{Tr}\Big[  u_{\mathbf{p},s, c} u^\dagger_{\mathbf{p},s, c}  \gamma^0\,  \hat{G}(\mathbf{x}_2,\mathbf{x}_1) \Big]\,  ,
\label{eq:Zqfinal}
\end{align} 
where $\int \mathcal{D} \Psi' \mathcal{D} \bar{\Psi}' = \int  d \Psi_1d \Psi_2\int_{\Psi_1}^{\Psi_2} \mathcal{D} \Psi \mathcal{D} \bar{\Psi}$ and $\hat{G}\equiv  ( i\slashed{D}[A]-m)^{-1}$ is the quark propagator. We used that $\hat{G}(\mathbf{x}_1,\mathbf{x}_2 ) \gamma^0 = \gamma^0 \hat{G}(\mathbf{x}_2,\mathbf{x}_1 ) $ at equal times $t_1=t_2$   in
the last equality of \Eq{eq:Zqfinal}. The trace in \Eq{eq:Zqfinal} and the indices $i,j$ are over spin and color. 

Employing the worldline representation of the fermion determinant, $\det( -i \hat{G}^{-1} )\ = \exp{(i\Gamma[A])} $, yields
\begin{align}\label{eq:effectiveactionwl}
\Gamma[&A] = i\text{Tr}\log(-i\hat{G}^{-1}) = \text{tr}_c\, i \int\limits_P \mathcal{D}x \mathcal{D}p \int\limits_{AP} \mathcal{D}\theta\mathcal{D}{\theta^*}\int \frac{\mathcal{D}\epsilon {\mathcal{D}\chi} }{\text{Vol}} 
\nonumber\\
&\times \exp\Big\{ i \int d\tau \, \dot{x}_\mu p^\mu  -\frac{i}{2} \dot{\theta}_i \theta_i^*+\frac{i}{2} {\theta}_i \dot{\theta}_i^*- H[x,p,\theta,\theta^*;A] \Big\}\,,
\end{align} 
where $i=1,2$ and the Hamiltonian given by
\begin{align}\label{eq:HamiltonianGaugefixed}
 H[x,p,\theta,\theta^*;A]&= \frac{\epsilon}{2}(P^2 +ig\psi^\mu  F_{\mu\nu}[A]\psi^\nu) - \frac{i\chi}{2}\,  P_\mu \psi^\mu \,,
\end{align}
with $\psi^0 = (\theta_1^*- \theta_1)/\sqrt{2}$, $\psi^3=(\theta_1^*+\theta_1)/\sqrt{2}$, $\psi^1=(\theta_2^*+\theta_2)/\sqrt{2}$ and $\psi^2=-i(\theta_2^*-\theta_2)/\sqrt{2}$ in Minkowski spacetime.  It is convenient to keep the $\theta_i , \theta_i^*$, instead of the Majorana representation $\psi^\mu$ when we give explicit expression of the valence quark initial density matrix  in Appendix \ref{app:densitymatrixproton}. 

Naive quantization of $\psi^0 \rightarrow \hat{\psi^0}$ as in \Eq{eq:WLHamiltonian} leads to states with indefinite metric. To consistently quantize the Dirac theory we therefore need to 
restrict the path integral measure $\mathcal{D}\psi \equiv i \mathcal{D}\theta\mathcal{D}{\theta^*}$ in \Eq{eq:effectiveactionwl} to the physical subspace of the Hamiltonian in \Eq{eq:HamiltonianGaugefixed}. The Dirac constraint defining this subspace is implemented via an (anticommuting) Lagrange multiplier variable $\chi(\tau)$ and the mass-shell constraint via the commuting Lagrange multiplier, the ``einbein" $\epsilon(\tau)$. Note that in this more general real-time formulation the $dT/T$ integral in \Eq{eq:effectiveaction1} is replaced by the integral over  $\epsilon(\tau)$. For details of this ``BRST fixing" of a gauge symmetry related to worldline reparametrization, we refer the reader to the discussion in \cite{Bastianelli:2006rx,Mueller:2019gjj}. In this formulation, ``$\text{Vol}$'' denotes the volume of the gauge group \cite{Bastianelli:2006rx}.  

The Dirac and mass shell constraints can be solved on the operator level to eliminate  $\hat{\psi}^0 $ and $\hat{p}^0 $. 
These are given by $\hat{\psi}^0 = \hat{p}^i \hat{\psi}^i/p^0$, where $p^0 = \pm |\mathbf{p}|$ (at leading order in the coupling $g$) acting on states of definite three-momentum $\mathbf{p}$. Solutions of the constraint equation at higher order in $g$ are given in section III of \cite{Mueller:2017arw}. 

Defining,
\begin{align}\label{eq:initialdensitymatrixsingleq}
 &\langle x_1,-\theta_1 | \hat{\rho}_q | x_2, \theta_2 \rangle \equiv \langle -\theta_1 |    u_{\mathbf{p},s} u^\dagger_{\mathbf{p},s}  \gamma^0  | \theta_2 \rangle  \, \Phi_c \Phi_c^\dagger
 \nonumber\\
&\times  2E_\mathbf{p}\, \delta(x^0_1(\tau=0)-t_0)\,  \delta(x^0_2(\tau=T)- t_0) \,e^{i p_\mu(x_1^\mu - x_2^\mu)}\,,
\end{align}
where $t_0\rightarrow -\infty$, and using Schwinger's proper time representation of the quark propagator in \Eq{eq:Zqfinal} we write~\footnote{Note the volume {Vol'} of the gauge group in \Eq{eq:fullthingSK} differs from that in \Eq{eq:effectiveactionwl}, due to the absence of zero modes present for periodic boundary conditions.}
\begin{align}
&2E_\mathbf{p} \int d^3\mathbf{x}_1d^3\mathbf{x}_2  \, e^{i\mathbf{p}(\mathbf{x}_1 - \mathbf{x}_2)}   \text{Tr }\Big[  u_{\mathbf{p},s, c} u^\dagger_{\mathbf{p},s, c} \gamma^0\,  \hat{G}(\mathbf{x}_2,\mathbf{x}_1) \Big]
  \nonumber\\
=& \text{tr}_c\, i \int d^4{x}_1d^4{x}_2 \int  d^2\theta_1d^2\theta_2 \langle x_1,-\theta_1 | \hat{\rho}_q | x_2, \theta_2 \rangle  
 \nonumber\\
 &\times\int\limits_{x_1}^{x_2} \mathcal{D}x \mathcal{D}p \int\limits_{\theta_1}^{\theta_2} \mathcal{D}\theta\mathcal{D}{\theta^*} \int\frac{{\mathcal D}\epsilon {\mathcal D}\chi}{\text{Vol}'}   \exp\Big\{ i \int d\tau \, \dot{x}_\mu p^\mu 
   \nonumber\\
   &   -\frac{i}{2} \dot{\theta}_i \theta_i^* +\frac{i}{2} {\theta}_i \dot{\theta}_i^* -H(x,p,\theta,\theta^*) \Big\}\,,\label{eq:fullthingSK}
\end{align}

With this and \Eq{eq:effectiveactionwl}, we can formally express the partition function of the proton with three valence quarks as
\begin{align}\label{eq:finalanswerSKproton}
&Z=\text{tr}_c\, i^3\int dA_1  dA_2\,    \int  \left[\prod\limits_{k=1}^3 d^4x_1^k d^4x_2^k d^2\theta_1^k d^2\theta_2^k  \right]
\nonumber\\
&\times  \langle A_1 | \hat{\rho}_A | A_2 \rangle \langle x_1, -\theta_1 | \hat{\rho}_V | x_2 ,\theta_2 \rangle 
\nonumber\\
&\times\int\limits_{A_1}^{A_2} \mathcal{D}A \left[   \prod\limits_{k=1}^3 \int\limits_{x^k_1}^{x^k_2} \mathcal{D} x^k \mathcal{D}p^k \int\limits_{\theta^k_1}^{\theta^k_2} 
\mathcal{D} \theta^k \mathcal{D}\theta^{*k} \frac{\mathcal{D}\epsilon^k{ \mathcal{D}\chi^k}}{\text{Vol}'}\right] 
\nonumber\\
&\times
\exp{ \Big\{ i S^{\text{YM}}_{\mathcal{C}}} + i \sum\limits_{k=1}^3 S_\mathcal{C}^k \Big\}\,  \exp{\Big\{ i \Gamma[A] \Big\}}\,,
\end{align}
where the fermion worldline action is
\begin{align}
S_\mathcal{C}^k \equiv \int d\tau\Big\{  \, \dot{x}^k_\mu p^{k,\mu} -\frac{i}{2} \dot{\theta}^k_i {\theta}_i^{*k} +\frac{i}{2} {\theta}^k_i {\dot{\theta}}_i^{*k} -H\Big\}\,,
\end{align}
and the Hamiltonian is given by \Eq{eq:HamiltonianGaugefixed}. Here  $i=1,2$ label the components of the Grassmann variables defined in \Eq{eq:coherentstatesdef}, while $k$ labels the valence quarks. The expression for the three valence quark
initial density matrix for this worldline path integral including spin, color and flavor is given in Appendix \ref{app:densitymatrixproton}. For the sake of a compact notation, we omit henceforth writing $\mathcal{D}\epsilon\mathcal{D}\chi $ explicitly and shall instead consider it to be part of the worldline path integral measure.
\begin{figure}[tb]
\begin{center}
\includegraphics[width=0.35\textwidth]{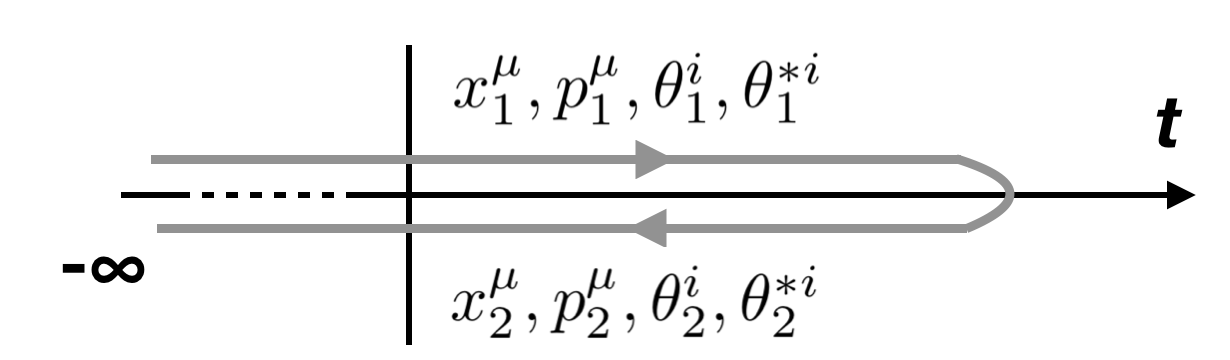}
\caption{The worldline path integrals for the quark effective action and propagator \Eqs{eq:effectiveactionwl}{eq:fullthingSK} have support on a Schwinger-Keldysh contour.
 \label{fig:SKcontour}}
\end{center}
\end{figure}

Upon (gauge-)fixing the worldline parametrization, by identifying worldline time $\tau\in[0,T]$ with physical time $x^0_{1/2}$ for upper and lower Keldysh contours,
\begin{align}\label{eq:wllightconegauge}
x^0_1\equiv x^0(\tau) &= t_0+\tau\,,  \qquad\qquad\quad\,\,  \tau \leq T/2\,,
\nonumber\\
x^0_2\equiv x^0(\tau)&=t_0+  (T-\tau),\qquad\; \; \;\tau > T/2\,,
\end{align}
 \Eq{eq:finalanswerSKproton} becomes a Schwinger-Keldysh path integral in physical time $x^0$ for the time evolution of bosonic and Grassmann worldline variables as well as Yang-Mills fields; See \Fig{fig:SKcontour} for illustration. In this coordinate-fixed formulation, we set the einbein parameter $\epsilon = 1/p^0$ and employ the temporal-axial gauge $A^0=0$.
 
The hadron tensor in \Eq{eq:Tmunu}  is defined by an ``in-in" matrix element of time-ordered electromagnetic currents. Introducing 
 (auxiliary) electromagnetic fields $a_\mu(x)$, we can relate this matrix element to the proton partition function \Eq{eq:finalanswerSKproton},
 \begin{align}
& \langle P,S| \mathbb{T} \, \hat{j}^\mu(\mathbf{x}) \hat{j}^\nu(0) | P,S\rangle = \text{Tr}\left[ \hat{\rho}_{P,S}\, \mathbb{T} \{ \hat{j}^\mu(\mathbf{x}) \hat{j}^\nu(0) \}\right]
 \nonumber\\
&  = \frac{\delta^2 Z }{i\delta a_{\mu}(z) \, i\delta a_\nu(0)}  \,.
 \end{align}
Using  the definition
 \begin{align}
&\left[ \prod\limits_{k=1}^3 \int\limits_{x^k_1}^{x^k_2} \mathcal{D} x^k \mathcal{D}p^k \int\limits_{\theta^k_1}^{\theta^k_2} 
\mathcal{D} \theta^k \mathcal{D}\theta^{*k} \right]
\exp{ \Big\{ i \sum\limits_{k=1}^3 S_\mathcal{C}^k \Big\}}
\nonumber\\
&\equiv\langle x_2,\theta_2 | \hat{\mathcal{U}}^{(3)}_{(-\infty,\infty)}[A ;a] \, \hat{\mathcal{U}}^{(3)}_{(\infty,-\infty)}[A;a]  | x_1,\theta_1 \rangle \,,
 \end{align}
 where $|x,\theta\rangle = \prod_{k=1}^3 |x^k,\theta^k\rangle$,
and the worldline time evolution operator for three valence quarks is
 \begin{align}
\hat{\mathcal{U}}^{(3)}_{(t,t')}[A;a]\equiv \exp{\Big\{-i \sum\limits_{k=1}^3\, \hat{H}^k[A;a](t-t')\Big\}}\,,
\end{align}
we can write the hadron tensor as
\begin{align}
&W^{\mu\nu}(q,P,S) = \frac{1}{\pi e^2} \text{Im} \, \text{tr}_c\, i^3 \int d^4z  e^{iq\cdot z} \int dA_1 dA_2
\nonumber\\
&\times \int   \left[\prod\limits_{k=1}^{3} d^4x_1^k d^4x_2^k d^2\theta_1^k d^2\theta_2^k  \right] \langle A_1 | \hat{\rho}_{\scriptscriptstyle{YM}} | A_2 \rangle
\nonumber\\ 
&\times  \langle x_1, -\theta_1 | \hat{\rho}_V | x_2 ,\theta_2 \rangle  \int \mathcal{D}A  \frac{\delta^2 }{i\delta a_{\mu}(z) \, i\delta a_\nu(0)}
\nonumber\\ 
&\times\Big[ \langle x_2,\theta_2 | \hat{\mathcal{U}}^{(3)}_{(-\infty,\infty)}[A ;a]  \hat{\mathcal{U}}^{(3)}_{(\infty,-\infty)}[A;a]  | x_1,\theta_1 \rangle 
\nonumber\\ 
&\times\exp{\{ iS_\mathcal{C}^{\text{YM}}[A]+i\Gamma[A;a]\}} \Big]\Big|_{a=0}\label{eq:WmunuQC}\,.
\end{align}
Here, $\Gamma[A;a]$ is given in \Eq{eq:effectiveactionwl}. The time ordering in this expression is such that one current operator insertion is on the upper (representing the amplitude), the other on the lower (representing the conjugate amplitude) Keldsyh contour~\cite{Gelis:2006yv}.

\begin{figure}[tb]
\includegraphics[width=0.27\textwidth]{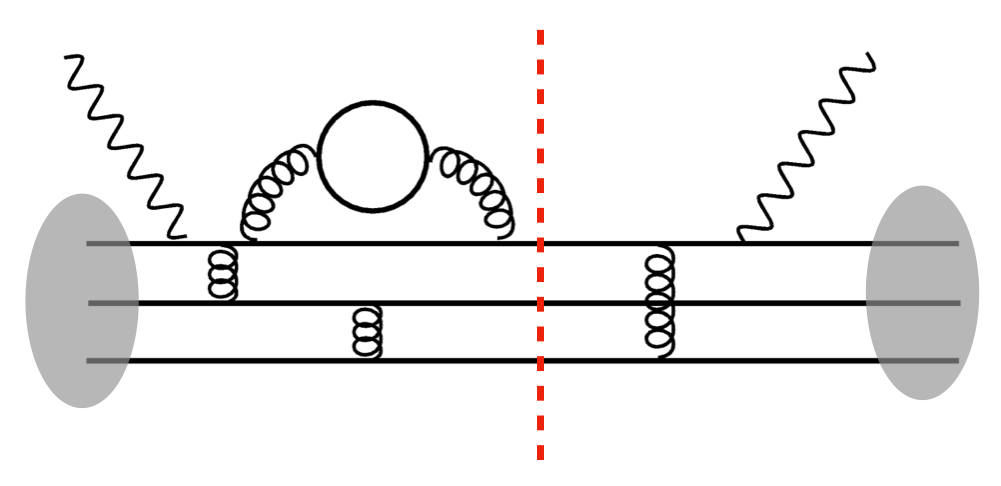}
\includegraphics[width=0.3\textwidth]{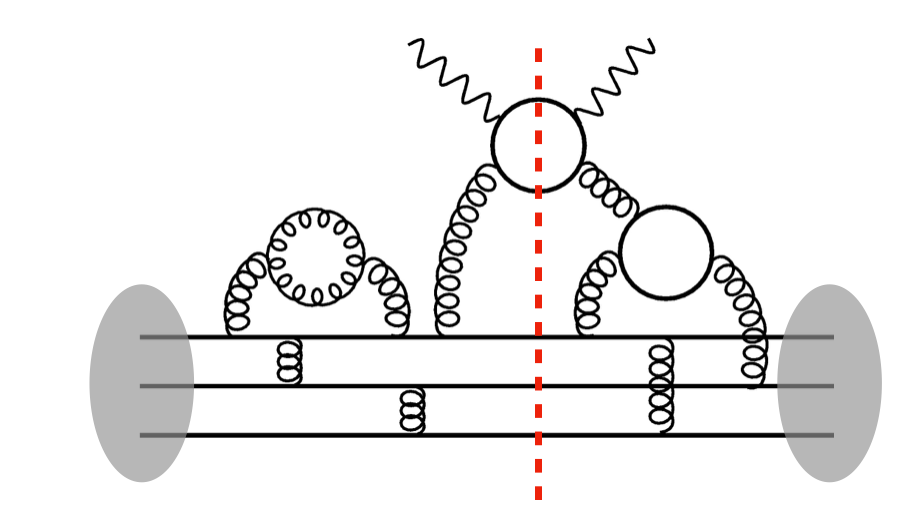}
\caption{Top: photon interaction with valence quarks. Bottom: Interaction with quark-antiquark pairs created from the color field of the proton.\label{fig:Wvis}. The red-dashed line
denotes the imaginary part taken via Cutkosky rules or equivalently the separation of the amplitude and the conjugate amplitude in this process.}
\end{figure}

The derivative in \Eq{eq:WmunuQC} yields two terms,
\begin{align}
&W^{\mu\nu}(q,P) = \frac{1}{\pi e^2} \text{Im}\text{ tr}_c \, i^3 \int d^4z  e^{iq\cdot z}\, \int dA_1 dA_2 \langle A_1 | \hat{\rho}_A | A_2 \rangle
\nonumber\\
&\times\int   \left[\prod\limits_{k=1}^{3} d^4x_1^k d^4x_2^k d^2\theta_1^k d^2\theta_2^k  \right] \langle x_1, -\theta_1 | \hat{\rho}_V | x_2 ,\theta_2 \rangle \int \mathcal{D}A
\nonumber\\ 
&\times \Big\{ \langle x_2,\theta_2 | \hat{\mathcal{U}}^{(3)}_{(-\infty,z)}    \hat{J}_{(3)}^{\mu}(z)  \hat{\mathcal{U}}^{(3)}_{(z,\infty)}
\hat{\mathcal{U}}^{(3)}_{(\infty,0)}  \hat{J}_{(3)}^{\nu}(0) \hat{\mathcal{U}}^{(3)}_{(0,-\infty)} | x_1,\theta_1\rangle
 \nonumber\\
& + \langle x_2,\theta_2 |\hat{\mathcal{U}}^{(3)}_{(-\infty,\infty)}  \hat{\mathcal{U}}^{(3)}_{(\infty,-\infty)}  | x_1,\theta_1 \rangle \, i\Gamma^{\mu\nu}[A](z,0) \Big\}
 \nonumber\\
 &\times \exp{\{ iS_\mathcal{C}^{\text{YM}}[A]+i\Gamma[A]\}}\,,\label{eq:Wmunudetails}
\end{align}
The first term is the valence quark contribution where the derivative acts on any of the valence quarks,
\begin{align}\label{eq:currentops}
&\frac{\delta }{i \delta a_\mu(z)}\mathcal{U}_{(\mp \infty,\pm\infty)}  \equiv \mathcal{U}_{(\mp \infty,z)}\hat{J}_{(3)}^\mu(z)\,    \mathcal{U}_{(z,\pm \infty)}\,,
\end{align} 
as depicted at the top of \Fig{fig:Wvis}.
Here, $\hat{J}_{(3)}^\mu(z) \equiv\sum_{k=1}^n \hat{j}_k^\mu (z)$ can be explicitly computed by varying the worldline Hamiltonian 
\begin{align}
\hat{j}_k^\mu(z)\equiv \frac{e}{p_k^0} \Big[\hat{P}_k^\mu + i \hat{\psi}_k^\nu\hat{\psi}_k^\mu q_\nu \Big] \delta^{(3)}(\mathbf{z}-\hat{\mathbf{x}}_k(z^0) )\,.
\end{align}

The second term in  \Eq{eq:Wmunudetails}, where the derivative acts on the exponential $\exp{(i\Gamma[A;a])}$, yields the photon polarization tensor,
\begin{align}\label{eq:poltensorgeneral}
&i\Gamma^{\mu\nu}[A](z,0) \equiv \text{tr}_c \, i \int d^4x \, d^2\theta \langle x,-\theta | \hat{\mathcal{U}}_{(-\infty,z)}    \hat{j}^\mu_{}(z)  \hat{\mathcal{U}}_{(z,\infty)}
\nonumber\\
&\times  \hat{\mathcal{U}}_{(\infty,0)}  \hat{j}^\nu(0) \hat{\mathcal{U}}_{(0,-\infty)} | x,\theta\rangle\,.
\end{align}
In the dipole picture, this term may be understood as the virtual photon fluctuating into a quark-antiquark pair which subsequently interacts with the color field of the target (see  bottom figure of \Fig{fig:Wvis}).
This term provides by far the dominant contribution to $F_2$ in the high energy limit of the CGC EFT with the first term suppressed by $x_{\rm Bj}$ as $x_{\rm Bj}\rightarrow 0$.

To bring \Eq{eq:Wmunudetails} into a form useful for quantum simulation, we insert a complete set of states into \Eq{eq:poltensorgeneral} and write
\begin{align}
i\Gamma^{\mu\nu}[&A](z,0) = \text{tr}_c \, i\int d^4x_1  d^4x_2 \int d^2\theta_1 d^2\theta_2 
\nonumber\\
& \times   \langle x_1,-\theta_1 | \mathbb{I}  | x_2,\theta_2 \rangle  \langle x_2,\theta_2| \hat{\mathcal{U}}_{(-\infty,z)}    \hat{j}^\mu(z)  \hat{\mathcal{U}}_{(z,\infty)}
\nonumber\\
& \times \hat{\mathcal{U}}_{(\infty,0)}  \hat{j}^\nu(0) \hat{\mathcal{U}}_{(0,-\infty)} | x_1,\theta_1\rangle\,.
\end{align}
We then perform a loop expansion of $\exp{(i\Gamma[A])} = \sum_{n=0}^\infty (i\Gamma[A])^n/n! $, allowing us to express the hadron tensor \Eq{eq:Wmunudetails} as 
\begin{align}
\label{eq:Wmunu-n}
W^{\mu\nu}(q,P,S) = \sum\limits_{n=0}^\infty \frac{1}{n!}W_{(n)}^{\mu\nu}(q,P,S)\,,
\end{align}
where the  $n$-quark loop contribution is given by
\begin{align}\label{eq:expansionFock}
&W_{(n)}^{\mu\nu} =\frac{1}{\pi e^2} \text{Im}\, \text{tr}_c \int d^4 z \, e^{iq\cdot z}    \int  \left[\prod\limits_{k=1}^{4+n} d^4x_1^k d^4x_2^k d^2\theta_1^k d^2\theta_2^k \right]
\nonumber\\
&\times i^{4+n} \int dA_1  dA_2\,  \langle A_1 | \hat{\rho}_A | A_2 \rangle   \langle x_1, -\theta_1 | \hat{\rho}_V \otimes \hat{\rho}_\Gamma \otimes \mathbb{I}_{n} | x_2 ,\theta_2 \rangle
\nonumber\\
&\times\langle x_2,\theta_2, A_2 | \hat{\mathbb{U}}_{(-\infty,z)}    \hat{J}^\mu_{(4)}(z)  \hat{\mathbb{U}}_{(z,\infty)}  
\nonumber\\
&\times \hat{\mathbb{U}}_{(\infty,0)}  \hat{J}_{(4)}^\nu(0) \hat{\mathbb{U}}_{(0,-\infty)} | x_1,\theta_1, A_1\rangle\,
\end{align}
where $|x,\theta,A\rangle=|x,\theta\rangle |A\rangle$. 
The worldline and Yang-Mills evolution operator is $\hat{\mathbb{U}}_{(t,t')}  \equiv \exp{\{-i  \hat{H}(t-t')\}}$
with $\hat{H}= \hat{H}_\text{YM}  +  \sum_{k=1}^{4+n}  \hat{H}^k $ being the sum of  the coordinate-fixed Hamiltonian of the $k$-th worldline $\hat{H}^k $
\begin{align}\label{eq:worldlineHamiltfixedmaintext}
\hat{H}^k &= \frac{1}{2p^0_k} \big( \hat{P}_k^2 + ig \hat{\psi}^{\mu}_k F_{\mu\nu}[{A}(\hat{x}_k)]  \hat{\psi}^{\nu}_k +ig \hat{\psi}^{\mu}_k F_{\mu\nu}[{a}(\hat{x}_k)]  \hat{\psi}^{\nu}_k \big),
\end{align}
and the Yang-Mills Hamiltonian in temporal-axial gauge,
\begin{align}
\hat{H}_\text{YM} = \int d^3x \, \frac{1}{2}[\hat{\mathbf{E}}^a(\mathbf{x})]^2 +  \frac{1}{2}[\hat{\mathbf{B}}^a(\mathbf{x})]^2,
\end{align}
where $\mathbf{\hat{E}}^a(\mathbf{x})$ and $\hat{B}^{i,a}(\mathbf{x})=\epsilon^{ijk} \hat{F}^{a,jk}(\mathbf{x})/2$ are the chromo-electric and chromo-magnetic field operators respectively.

To summarize, the hadron tensor \Eq{eq:Tmunu} in its most general form can be expressed as as a hybrid Yang-Mills and quark worldline path integral 
\begin{align}\label{eq:masterformula}
&W^{\mu\nu} =  \frac{1}{\pi e^2}\text{Im}\,  \int d^4 z \, e^{iq\cdot z}\sum\limits_{n=0}^\infty  \frac{i^{n+4}}{n!} \int  \Big[\prod\limits_{k=1}^{n+4} d^4x_1^k d^4x_2^k 
\nonumber\\
&\times d^2\theta_1^k d^2\theta_2^k  \Big]
  \int dA_1  dA_2\;  \,  \text{tr}_c\;
\langle  x_1, -\theta_1, A_1 | \hat{\rho}_\text{init} | x_2 ,\theta_2, A_2 \rangle
\nonumber\\
&\times  \langle x_2,\theta_2, A_2 |   \hat{\mathbb{U}}_{(-\infty,z)} \hat{J}^\mu_{(4)}(z)  \hat{\mathbb{U}}_{(z,\infty)}  
\nonumber\\
&\times\hat{\mathbb{U}}_{(\infty,0)}  \hat{J}_{(4)}^\nu(0) \hat{\mathbb{U}}_{(0,-\infty)} | x_1,\theta_1, A_1\rangle\,,
\end{align}
It is important to note that $\hat{p}^0$ and $\hat{\psi}^0$ here are not dynamical operators and are removed from the physical Hilbert space by Dirac and mass-shell constraints, as discussed previously. 

\Eq{eq:masterformula} is the master formula for computing structure functions as the initial value problem  $\partial_t \hat{\rho} = -i[\hat{H}, \hat{\rho}]$ with the initial condition $\hat{\rho}_\text{init}$ followed by the measurement of electromagnetic worldline current operators. As discussed previously, the initial density matrix in this coherent state basis at past infinity $\hat{\rho}_{\text{init}} = \hat{\rho}_\text{YM} \otimes \hat{\rho}_{f} \otimes \hat{\rho}_\Gamma \otimes \mathbb{I}_{n}$. Here $ \hat{\rho}_\text{YM}=|0\rangle \langle 0| $ is the noninteracting Yang-Mills vacuum, $\hat{\rho}_{f}$ contains initial conditions for $k=1,2,3$ valence quarks, $k=4$ denotes the quark-antiquark dipole $ \hat{\rho}_\Gamma=\mathbb{I}$ in the polarization tensor $\Gamma^{\mu\nu}$, and $\mathbb{I}_{n}$ is a unit matrix representing the other $k=5,\dots,n$ ``sea-quark" Fock-states. 

\begin{figure}[tb]
\includegraphics[width=0.25\textwidth]{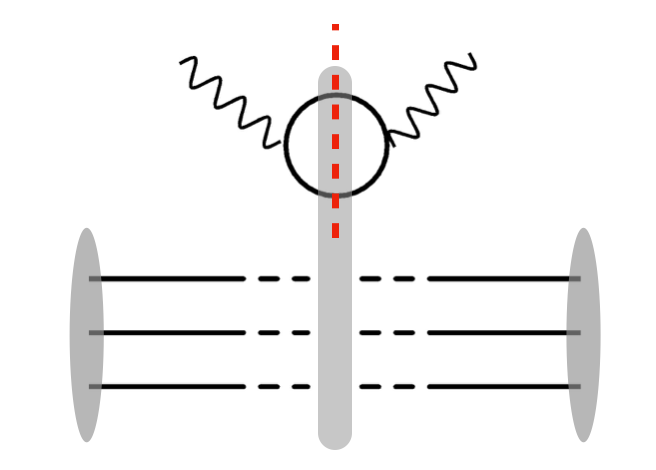}
\caption{High energy shock wave limit of the photon-proton interaction.\label{fig:CGCvis}}
\label{fig:f3}
\end{figure}

The simpler expression for $W^{\mu\nu}$ in \Eq{eq:hadrontensor} is obtained from \Eq{eq:masterformula} because of the separation of time scales between large and small $x_{\rm Bj}$ modes we alluded to previously. This limit is illustrated in \Fig{fig:f3}. Valence quarks ($k=1,2,3$) and large $x_{\rm Bj}$ partons become quasi-static color sources and therefore the tensor product representing their density matrix can be replaced with the weight functional $W[\rho]$~\cite{Jeon:2004rk}. Further, in  \Eq{eq:hadrontensor}, only the polarization tensor ($k=4$), representing the virtual photon splitting into a quark-antiquark pair is computed explicitly and the path integral over gauge fields is greatly simplified in the CGC EFT by performing the weak coupling expansion around  $A_{\rm cl}$. 

%

\section{Single particle digitization strategy}

The scheme we outlined thus far, of expressing the fermion sector of the QCD path integral via quantum mechanical ``worldlines'', may serve as a  template for a novel digitization strategy. We  illustrated here only the simplest part of such program, the treatment of internal symmetries, explicitly in terms of circuits. However the central result of this work is the general path integral formulation in \Eq{eq:masterformula}, from which a Hamiltonian time evolution scheme can be derived. We have yet to prove the practicality of our approach for actual computation, using 
quantum hardware resources that are realistically available. Specifically, for any significant extension beyond the simplest toy problem discussed here, it is essential that the bosonic worldline variables $\hat{x}$ and $\hat{p}$ be quantum simulated. We will outline below first steps in this direction by considering the qubit digitization of single (position and momentum) (quasi-)particle Hilbert spaces in scalar quantum field theory (QFT); this is work in progress that will be reported on in a follow-up paper~\cite{NAR}.

We propose the use of a basis of $M$ relativistic `worldline'/single particle states $\mathcal{H}=\bigotimes_{i=1}^M \mathcal{H}_{i}$, where $\mathcal{H}_i = \{ | \mathbf{p} \rangle^{(i)}\},\{ | \mathbf{x} \rangle^{(i)}\}$ are $N^d$ discretized momentum and position eigenstates $\mathbf{p} \equiv \mathbf{p}_n, \mathbf{x} \equiv \mathbf{x}_m$  ($n,m\in \mathbb{Z}$)
that are binary-encoded in a sequence of $d \log_2(N)$ spin/qubits per particle register $i$ (plus  additional auxiliary qubit(s) for their occupation number/(fermion-)parity). Here $d=3$ denotes the dimension.  We emphasize that this basis should not be confused with the usual harmonic oscillator basis of Fock operators in momentum space.

Implementing this single particle/worldline digitization strategy requires  order  $O(M\text{log}_2(N^d))$ qubits, where $M$ is proportional to the number of external particles and loops, typically of order $M\sim O(10-50)$. Reflecting the large dynamical range of momenta probed  in high energy scattering experiments, typically $N^d\sim O(\text{100}^3)$. 
A more conventional digitization, e.g. constructed from local Hilbert spaces
of Heisenberg (lattice) field operators $\{ | \phi_\mathbf{x}  \rangle\}, \, \{ | \pi_\mathbf{x}  \rangle\}$ would require roughly $O(N^d \log(N_{\rm loc}))$ qubits, where $N_{\rm loc}$ is the size of the local field operator space~\cite{Jordan:2011ci,Jordan:2011ne,klco2019digitization}. 
In this comparison, our strategy is well suited for high energy scattering, but will fare worse for systems with very large occupation numbers.
 
To prepare a scattering `experiment', we create an initial state of $n$ (and $M$-$n$ non-occupied) non-interacting wave packets efficiently by a variant of (pair-wise) Bell-entanglement of aforementioned $n$ occupied and $M-n$ non-occupied single particle states $\{| \mathbf{p} \rangle^{(i)}\}$. This is very efficient in our basis.  We then use a Trotter scheme to perform unitary time evolution. The kinetic part of the Hamiltonian  is diagonal in our basis (of Bose- or Fermi-symmetrized single particle states) \cite{zalka1998efficient}.  Moreover one can directly work with the continuum dispersion relation instead of the lattice one\footnote{In the lattice digitization of Heisenberg field operators, the dispersion relation depends on the discretization of the kinetic term in lattice position space~\cite{klco2019digitization}. It can be improved by making this term more and more non-local.}, up to the largest $|\mathbf{p}|$ that can be realized for a given qubit digitization.

Interaction terms are highly non-local in the $\{| \mathbf{p }\rangle^{(i)}\}$ basis; however, we can transform to the $\{| \mathbf{x }\rangle^{(i)}\}$ basis to compute them. A difficulty is that $| \mathbf{p }\rangle^{(i)}$ and $| \mathbf{x}\rangle^{(i)}$ are not simply Fourier conjugates in a relativistic theory,  but are connected by a quantum Fourier transformation (which is particle-local) and a variant of a (`inter-mode') squeezing transformation~\footnote{In our case, this transformation is related, but distinct from the bosonic ``intra-mode" squeezing used in quantum optics~\cite{walls1983squeezed,wu1986generation}.}
(which in our basis is two-particle-local). In the resulting $| \mathbf{x}\rangle^{(i)}$ basis, interaction terms (for instance, the $\phi^4$ interaction term in a scalar field theory) involve (Bose- or Fermi-symmetrized) interactions between all particle sectors but are few-particle local (four in the case of $\phi^4$).  
To relate our procedure to an actual cross section, we measure non-interacting, asymptotic states  in the  basis $\{| \mathbf{p }\rangle^{(i)}\}$, following~\cite{Jordan:2011ci,Jordan:2011ne}. 

We will report on this strategy in a forthcoming paper~\cite{NAR}, including also a detailed resources analysis in particular for 
the squeezing and interaction part. We note that, for $d=1$ a simplified toy computation with few particles and on small systems requires only
tens of qubits, and thus may allow an error analysis, at least for some elements of the algorithm. We plan to explore whether this approach is competitive to that of~\cite{Jordan:2011ci,Jordan:2011ne}, in particular for scattering problems which typically have a few particles but a large position (and momentum) space volume.

\section{Outlook: Expanding in scope}
We have focused thus far on laying the conceptual foundations of our hybrid approach ``bottom-up'' for the physical system of interest (instead of a toy model); 
the high energy `'Regge'  limit is likely a fertile starting point for this approach.  This hybrid framework may also allow one  to quantum compute not just structure functions but in principle multi-leg and multi-loop scattering amplitudes~\footnote{We note that worldline methods have been extensively used in such computations~\cite{Bern:1991aq,Strassler:1992zr,Bern:1994cg,Schmidt:1996ih,Laenen:2008gt,Gardi:2010rn}; it would be interesting to explore computing the nontrivial color and spinor traces therein utilizing the worldline quantum circuits discussed in Appendix~\ref{app:circuit}.}. The diagrammatic expansion of Feynman amplitudes in perturbation theory exhibit factorial growth in the number of diagrams at each loop order~\cite{Jordan:2011ne,Jordan:2011ci}. In sharp contrast, such computations on a quantum computer with fermionic and bosonic worldline variables would require resources that only scale polynomially with the number of particles~\footnote{An important issue to address in this context is the renormalization of infinities at each loop order. These have been discussed extensively in \cite{Sato:1998sf,Sato:2000cr,Pawlowski:2008xh,Bastianelli:2013pta,Magnea:2015fsa}.}.

Extending our toy problem systematically within the CGC EFT is manageable because one would, as a first step, gauge covariantly couple the $|\{ | \mathbf{x} \rangle^{(i)}\}$ basis to the Yang-Mills part, which is treated
classically.
For the full non-perturbative problem \Eq{eq:masterformula}, the dominant part of resources would be used to realize the Yang-Mills Hilbert space, using the Kogut-Susskind lattice Hamiltonian approach \cite{Kogut:1974ag}.  This is a difficult problem and its implementation on analog and digital quantum devices is deservedly a subject of much attention~\cite{Wiese:2013uua,martinez2016real,klco2018quantum,Alexandru:2019nsa}, prominent examples being quantum link models~\cite{Wiese:2013uua} and the matrix product state formalism~\cite{Byrnes:2002nv,sugihara2005matrix,Cirac:2009zz,Buyens:2013yza,Banuls:2018jag,tagliacozzo2011entanglement}.  Some related recent proposals are discussed in \cite{Anishetty:2009nh,Banerjee:2012xg,Zohar:2012xf,zohar2015formulation}. 

Preparing the projectile and target state is a nontrivial problem requiring a large number of time evolution steps; for example, in a Trotter scheme, this would imply a large number of gate operations. Moreover simulation errors drive the system from the physical part of the Hilbert, requiring high fidelity of the gate operations.  It is conceivable that different approaches \cite{Anishetty:2009nh,Banerjee:2012xg,Zohar:2012xf,zohar2015formulation} will have different sensitivity to errors and should be investigated further. An advantage of the high energy limit is that the initial proton state at small $x_{\rm Bj}$
is not in the ground state but in a highly excited state, implying shorter preparation time.

We note finally that this phase space worldline formalism permits a semi-classical Moyal expansion~\cite{Moyal:1949sk} to construct Wigner functions~\cite{Mueller:2019gjj}.  These are accessible in DIS~\cite{Burkardt:2002hr,Belitsky:2003nz} and therefore allow one to probe systematically, at higher orders in $\hbar$, parton entanglement in QCD at high energies~\cite{Kharzeev:2017qzs,Berges:2017zws,Beane:2019loz}.

\section*{Acknowledgments.}
We thank P. Bedaque, J. Berges, M. Creutz, P. Hauke, H. Lamm, P. Love, M. McGuigan, P. Petreczky, R. Pisarski, M. Savage and T. Zache for discussions on quantum information science. We would like to thank the Institute for Nuclear Theory at the University of Washington and the ITP Heidelberg for their kind hospitality during the completion of this work. 
The authors are supported by the U.S. Department of Energy, Office of Science, Office of Nuclear Physics, under contract No. DE- SC0012704 and by U.S. DOE grant DE-SC0004286 (AT), within the framework of the Beam Energy Scan Theory (BEST) Topical Collaboration and the Topical Collaboration for the Coordinated Theoretical Approach to Transverse Momentum Dependent Hadron Structure in QCD (TMD Collaboration). NM is funded by the Deutsche Forschungsgemeinschaft (DFG, German Research Foundation) - Project 404640738. %
%
%
%
%
%
\bibliographystyle{apsrev4-1} 

\appendix

\section{Worldline representation of the proton's spin, flavor and color valence structure}\label{app:densitymatrixproton}
In this Appendix, we discuss the representation of the proton's initial density matrix composed of the valence quarks, including color, spin and flavor. The valence wave function
of a proton polarized along its momentum ($S=1/2$) is a direct product of the a color-singlet and a combined spin/flavor part $\hat{\Phi}_{P,S} |0\rangle = | \text{color} \rangle \otimes | \text{spin/flavor}\rangle$~\cite{Griffiths:1987tj},
\begin{align}\label{eq:matrixflavorspin}
| \text{spin/flavor}\rangle =& \frac{\sqrt{2}}{3} | u(\uparrow)u(\uparrow)d(\downarrow)\rangle - \frac{{1}}{3\sqrt{2}}  | u(\uparrow)u(\downarrow)d(\uparrow) \rangle 
\nonumber\\
&- \frac{{1}}{3\sqrt{2}}  | u(\downarrow)u(\uparrow)d(\uparrow)\rangle + \text{(permutations)}\,,
\end{align}
where $\uparrow(\downarrow)$ represents a quark with spin $s=1/2$ $(-1/2)$ while
\begin{align}\label{eq:colormatrixdensity}
| \text{color} \rangle = \frac{1}{\sqrt{6}}\sum\limits_{ijk\in{(r,g,b)}} \epsilon^{ijk}| i,j,k\rangle \,.
\end{align}

We outline first the worldline representation of the spin part. The single quark spin density matrix with spin $s=\pm1/2$, assuming a right-moving valence quark with large ${p}^+\approx \sqrt{2} |\mathbf{p}|$, can be written as 
\begin{align}\label{eq:densitymatrix}
u_{\mathbf{p},s} u^\dagger_{\mathbf{p},s}  &=  \frac{1}{2}\left( 1+2s \gamma_5 \right) {p}^+ \gamma^- \gamma^0\,.
\end{align}
Using the representation of the Clifford algebra of the Lorentz group and the representation of the chirality matrix in terms of fermion creation and annihilation operators $\hat{b}_i^\dagger, \hat{b}_i$,
\begin{align}
\gamma_5 = -(-1)^{\sum_{j=1}^2\hat{b}^\dagger_j \hat{b}_j} = -\prod\limits_{j=1}^2(1-2\hat{b}^\dagger_j \hat{b}_j )\,,
\end{align}
we obtain
\begin{align}\label{eq:matrixspin}
u_{\mathbf{p},s} u^\dagger_{\mathbf{p},s} &=   |\mathbf{p}| \Big[ 1- 2s  \prod_{j=1}^2(1-2\hat{b}_j^\dagger \hat{b}_j) \Big] (1-\hat{b}^\dagger_1\hat{b}_1)\,.
\end{align}
Using the Jordan Wigner transformation, this may be written as
\begin{align}
u_{\mathbf{p},s} u^\dagger_{\mathbf{p},s} =\frac{ |\mathbf{p}| }{2}\big\{ \mathbb{I}\otimes \mathbb{I}+ \sigma^z\otimes\mathbb{I} - 2s\, \mathbb{I}\otimes \sigma^z - 2 s\, \sigma^z \otimes \sigma^z\big\}\,.
\end{align}%

To write the color structure of the worldline density matrix and the  effective action, we first introduce the representation of the color trace $\text{tr}_c$ of an operator $O$ in terms of Grassmann coherent states,
\begin{align}
\text{tr}_c O = \int d^3 \lambda \langle -\lambda | O | \lambda \rangle\,.
\end{align}
More generally, the trace over  a path ordered color matrix exponential in this formalism is~\cite{Barducci:1980xk,DHoker:1995uyv,DHoker:1995aat}  
\begin{align}\label{eq:Lagrange}
&\text{tr}_c \mathcal{P}\exp{\left[i \int_0^T d\tau M(\tau)\right] } = \int \mathcal{D}\phi\int \mathcal{D}\lambda^\dagger \mathcal{D}\lambda\, \nonumber\\
&\times e^{i\phi (\lambda^\dagger \lambda +\frac{n}{2}-1)}\exp{\Big[ i \int_0^T d\tau ( i\lambda^\dagger \frac{d\lambda}{d\tau} - \lambda^\dagger M \lambda ) \Big]}\,,
\end{align}
where the Grassmann variables  $\lambda_i ,\lambda^*_i$, with $i=1,2,3$ being Wigner-Weyl symbols of fermonic operators $\hat{c}_i, \hat{c}_i^\dagger$,
\begin{align}\label{eq:creationcolor}
\hat{c}_i | \lambda \rangle = \lambda_i | \lambda \rangle\,, \qquad  \hat{c}_i^\dagger  | \lambda^* \rangle = \lambda_i^* | \lambda^* \rangle\,,
\end{align}
similar to \Eqs{eq:representationGamnma}{eq:Majorana} for spin. In \Eq{eq:Lagrange}, $\phi$ is the Lagrange multiplier implementing the constraint restricting the fermion creation and annihilation operators to act on a finite dimensional representation of $SU(3)$.
The color-matrix valued coordinate fixed worldline Hamiltonian in \Eq{eq:worldlineHamiltfixedmaintext} may therefore be generalized to
\begin{align}\label{eq:colorHamilt}
&\hat{H}^k = \frac{1}{2p^0_k} \big( \hat{P}_k^2 + ig \hat{\psi}^{\mu}_k \hat{c}^{k\dagger}_{b} F^{a}_{\mu\nu}[A(\hat{x}_k)]\,t^a_{bc}  \hat{c}_c^k   \hat{\psi}^{\nu}_k 
\nonumber\\
&+ig \hat{\psi}^{\mu}_k F_{\mu\nu}[{a}(\hat{x}_k)]  \hat{\psi}^{\nu}_k \big),
\end{align}
Here, $\hat{P}_\mu=p_\mu-ig \hat{c}^\dagger_b A_\mu^a(x) t^a_{bc} \hat{c}_c-iea_\mu(x)$ and $n=3$ is the dimension of the matrix $M(\tau)$.

A single quark color matrix can be written in terms of the unit matrix $\mathbb{I}_3$ and the $SU(3)$ generators  $t^a$ (in fundamental representation) as 
$\hat{\rho}_{\text{color}}= \mathbb{I}_3/3 + \sum_{a=1}^8\,\phi^a\, t^a$, where $\phi^a$ are c-number coefficients. In the worldline formulation, where color traces are expressed through Grassmann coherent states, this may equivalently written as
\begin{align}\label{eq:colordensity}
\hat{\rho}_\text{color}=  1+ \frac{2  d^{abc} }{A_Rd^2}t^a_{ij} t^b_{kl}t^c_{mn} \, \hat{c}_i^\dagger \hat{c}_j\hat{c}_k^\dagger \hat{c}_l \hat{c}_m^\dagger \hat{c}_n  + 2 \phi ^a\,  t^a_{ij} \hat{c}_i^\dagger \hat{c}_j\,,
\end{align}
as derived in \cite{Mueller:2019gjj}.

In the (r,g,b) basis the coefficients $\phi^a$ are
\begin{align}
\phi^a =  \Bigg\{ 
\begin{array}{lr}
\delta^{a3} + \frac{1}{\sqrt{3}}\delta^{a8} &  \text{ (red)} \\
-\delta^{a3} + \frac{1}{\sqrt{3}}\delta^{a8} &  \text{ (green)}\\
- \frac{1}{\sqrt{3}}\delta^{a8} &  \text{ (blue)}
\end{array}\,.\label{eq:coefficients}
\end{align}
The color-flavor-spin density matrix of a baryon, containing three valence quarks is written as the product 
of a symmetric spin-flavor \Eq{eq:matrixflavorspin} and an antisymmetric color-singlet part \Eq{eq:colormatrixdensity}. Employing the Jordan-Wigner transformation, as in the spinor case, the color density matrix may be written as a three-qubit quantum circuit.

\section{Quantum Circuit for the worldline computation of $F_2$}
\label{app:circuit}
In this Appendix, we will present details of the quantum circuits required for the worldline computation of $F_2$ in \Eq{eq:F2}. To compute the trace in \Eq{eq:F2} we employ the circuit in \Eq{eq;tracecircuit}. Here the $n=2$ circuit qubits are initially in a mixed state with density matrix $\rho_2 = \mathbb{I}_2/2^2$, while an additional control qubit (in a pure state 
$|0\rangle \langle 0 |$) is used. The combined density matrix of control and circuit qubits after employing \Eq{eq;tracecircuit} is
\begin{align}\label{eq:combineddensitymatrix}
\rho_{2+c}=\frac{1}{2^3}
\begin{pmatrix}
\mathbb{I}_2 & \Omega_{L,T}^\dagger \\
\Omega_{L,T}&\mathbb{I}_2 
\end{pmatrix}\,,
\end{align}
so that measurement of $\sigma^x$ and $\sigma^y$ on the control qubit yields
\begin{align}\label{eq:traceapp}
\langle \sigma^x \rangle &= \text{ Tr} [\sigma^x \,\hat{\rho}] = \frac{1}{2^2}\text{Re}[ \text{Tr }\Omega_{L,T} ]\,,\nonumber\\
\langle \sigma^y \rangle &= \text{ Tr} [\sigma^y \,\hat{\rho}]  = -\frac{1}{2^2}\text{Im}[ \text{Tr }\Omega_{L,T} ]\,.
\end{align}
A crucial ingredient in \Eq{eq;tracecircuit} is the controlled gate $C(\Omega_{L,T})=\begin{pmatrix}
\mathbb{I} & 0 \\
0 & \Omega_{L,T}
\end{pmatrix}$. Since $\Omega_{L,T}$ is decomposed into more fundamental gates $\Omega_{L,T} = \prod_i G^i$, where the $G^i$ stand for the Hadamard gate
$H=\frac{1}{\sqrt{2}}\begin{pmatrix}
1 & 1 \\
1 & -1
\end{pmatrix}$ and the phase gate $S=\begin{pmatrix}
1 & 0 \\
0 & i
\end{pmatrix}$, we can construct  $C(\Omega_{L,T})$ from the controlled gates of its constituents~\cite{NielsenChuang},
\begin{align}\label{eq:decompositioncontroll}
C(\Omega_{L,T})  =\prod_i  C(G^i)\,.
\end{align}
This allows us to write the control circuit of \Eq{eq:photonvertex}. 

We can also use the fact that $C(\sigma^z)$ is a standard gate available on present and future hardware to write $C(\sigma^x) $ and $C(\sigma^y)$,
\begin{align} \label{eq;controlledxy}
&\Qcircuit @C=.5em @R=0em @!R {
& \ctrl{1} & \qw  & & & \ctrl{1} &  \ctrl{1} &  \ctrl{1} & \qw \\
& \gate{\sigma^x}& \qw  &  \push{\rule{.3em}{0em}=\rule{.3em}{0em}} & & \gate{H}  & \gate{\sigma^z} &  \gate{H} & \qw
}\,,\\
&\Qcircuit @C=.5em @R=0em @!R {
& \ctrl{1} & \qw  & & & \ctrl{1} &  \ctrl{1}&   \ctrl{1} & \qw \\
& \gate{\sigma^y}& \qw  &  \push{\rule{.3em}{0em}=\rule{.3em}{0em}} && \gate{SH}  & \gate{\sigma^z} &  \gate{HS^\dagger}  & \qw
}\,,
\end{align}
using 
\begin{align}\label{eq:standarident}
S\,H\, \sigma^z \, H S^\dagger &= \sigma^y\,\\
H\, \sigma^z \, H & =\sigma^x\,,
\end{align}
where $\Qcircuit @C=.2em @R=.2em{&\gate{SH} &  \push{\rule{.1em}{0em}=\rule{.1em}{0em}} & \gate{S} & \gate{H} & \qw }$. Implementations of controlled Hadamard- and phase-gates can be found for example in \cite{NielsenChuang,IBMQHaddamard}.

\bibliography{references}

\end{document}